\documentclass[journal,10pt,twocolumn,twoside]{IEEEtran}

\usepackage{cite}
\usepackage{amsmath,amssymb,amsfonts}
\usepackage{algorithmic}
\usepackage{graphicx}
\usepackage{subfigure}
\usepackage{textcomp}
\usepackage{xcolor}
\usepackage{subfigure}
\usepackage{graphicx}
\usepackage{caption2}

\newtheorem{lemma}{Lemma}

\newtheorem{proposition}{Proposition}

\newtheorem{theorem}{Theorem}
\newtheorem{remark}{Remark}

\ifCLASSINFOpdf

\else

\fi

\begin{document}

\title{Designing Unimodular Waveforms for MIMO Radar Based on Manifold Optimization Method}

\author{Xuyang Zhao,~\IEEEmembership{Student Member,~IEEE,}
        Jiangtao Wang,~\IEEEmembership{Member,~IEEE,}
        Shihao Yan,~\IEEEmembership{Senior Member,~IEEE,}\\
        and Yongchao Wang,~\IEEEmembership{Senior Member,~IEEE}}

\maketitle

\begin{abstract}
In this paper, we design unimodular waveforms with good correlation properties for multi-input multi-output (MIMO) radar systems. Specifically, first, we analyze the geometric properties of the unimodular constraint in the fourth-order polynomial minimization problem using Riemannian geometry theory. By embedding it into the search space, we transform the original non-convex optimization problem into an unconstrained problem on a Riemannian manifold. Then, we construct the manifold corresponding to the search space and the operators required for the customized optimization algorithm. Second, we develop a customized low-complexity unimodular manifold gradient descent (UM-GD) algorithm on the constructed manifold to solve the optimization problem in the normal-scale case, and propose its acceleration version unimodular manifold accelerated gradient descent (UM-AGD) algorithm, to speed up the convergence. In the large-scale case, we transform the objective function into the form of a summation of a large but finite number of loss functions and develop a customized unimodular manifold stochastic variance reduced gradient (UM-SVRG) algorithm to solve this problem. Compared to the existing bechmark method, which has a computational complexity of roughly $\mathcal{O}(M^4+3M^2N+3|\mathcal{D}| |\hat{\Theta}|^2MN)$, UM-SVRG algorithm effectively reduces the computational complexity of each iteration to roughly $\mathcal{O}(|\mathcal{D}||\hat{\Theta}|^2MN)$. Thirdly, we provide theoretical guarantees of convergence for both the UM-GD and UM-SVRG algorithms through appropriate parameter selection, and prove that the proposed algorithms can converge to a stationary point. Finally, numerical examples demonstrate the effectiveness of the proposed UM-GD and UM-SVRG algorithms.
\end{abstract}


\IEEEpeerreviewmaketitle

\section{Introduction}
\IEEEPARstart{W}{\MakeLowercase a}veform design is a cornerstone of multiple-input multiple-output (MIMO) radar technology, directly affecting system performance in terms of detection, resolution, and interference suppression\cite{Li-Jian}. Through waveform design, MIMO radar can generate transmitting beams with specific shapes and directions, concentrating energy in areas of interest to improve target detection performance and reduce energy leakage in non-target regions\cite{Phased-MIMO}. Additionally, by optimizing auto-correlation and cross-correlation properties of the waveform,  clutter and interference can be effectively suppressed, especially in complex environments with multipath propagations and strong clutters\cite{colocated-antennas}. Unimodular waveforms, which maintain consistent amplitude to enhance power efficiency, are widely used in MIMO radar systems. Thus, unimodular waveform design has become a key technical issue in MIMO radar systems\cite{MIMO-Beampattern}--\cite{residual-network}.

MIMO radar waveform design mainly focuses on the transmit beampattern design, waveform synthesis, and waveform design under various application constraints\cite{Peak-sidelobe}--\cite{J-Liang}.
The authors in \cite{PSL} used semidefinite programming (SDP) to match the waveform covariance matrix to the desired beampattern under unimodular constraint.
The authors in \cite{Aldayel} and \cite{Aldayel-TSP} employed continuous convex relaxation techniques to address non-convex quadratic equality constraints in unimodular waveform design. In \cite{Z-Cheng}, a double cyclic alternating direction method of multipliers (D-ADMM) algorithm was proposed to solve the beampattern design problem. The authors in \cite{CA} introduced a cyclic algorithm (CA) for integrated beampattern design, considering the correlation properties within a certain region. The authors in \cite{L-BFGS} modeled the waveform design problem as a quartic polynomial minimization problem with unimodular constraint and solved it by quasi-Newton method L-BFGS. In \cite{consensus-ADMM}, the authors proposed the consensus alternating direction method of multipliers (consensus-ADMM) algorithm and its heuristic version, the consensus-ADMM-AGD algorithm based on accelerated gradient descent (AGD), to solve the quartic polynomial minimization problem with unimodular constraint. In the aforementioned work, the handling of the unimodular constraint can be divided into two categories. The first category involves approximating or relaxing the unimodular constraint, where the design is done under approximate constraint and then the obtained solution is mapped to the constraint. The second category directly enforces the unimodular constraint, which offers better performance compared to the approximating methods but involves more complex computational processes.

In recent years, Riemannian manifold optimization has attracted much attention due to its ability to utilize the geometric structure of constraint. By embedding non-convex constraints into the search space, the original non-convex optimization problem is transformed into an unconstrained problem on the manifold, which improves the efficiency of obtaining the solution. In \cite{complex-circle}, the authors proposed a new projection, descent, and retraction (PDR) update strategy for solving the MIMO beampattern design problem on the complex circle manifold. In \cite{Li-Jie}, the authors developed a Riemannian gradient descent algorithm and Riemannian trust-region algorithm on the manifold to solve the joint design problem of transmit sequences and receive filter for airborne MIMO radar systems. The authors in \cite{Zhong-kai}--\cite{Zhong-ICASSP} proposed Riemannian conjugate gradient algorithm to solve the beampattern matching problem under different forms of constant modulus constraint. It is important to note that the aforementioned works only focused on the beampattern matching problem, which is different from the problem considered in this work, and therefore the form of the objective function is also different. Moreover, existing methods transform the original problem onto the complex circle manifold for solving, and require the optimization variables to be transformed into vector form, which leads to higher complexities. Additionally, in the works \cite{Zhong-kai}--\cite{Zhong-ICASSP}, the descent direction used is the Polak-Ribiere conjugate gradient descent direction and such methods lack theoretical convergence guarantees \cite{manifold-optimization-absil}, making them to some extent heuristic algorithms.

With the development of MIMO radar systems towards large-scale applications, reducing the computational complexity of waveform design has become an urgent challenge. This requires the development of efficient algorithms that can achieve superior waveform performance while maintaining computational feasibility in real-time applications. Currently, the consensus-ADMM algorithm proposed in \cite{consensus-ADMM} is suitable for large-scale MIMO radar systems because it can reformulate the original optimization problem into a consensus-like optimization model and decompose it into multiple subproblems for solving. However, it still suffers from high computational complexity, and the best-performing consensus-ADMM-AGD algorithm proposed in this work is a heuristic algorithm without convergence guarantees. Therefore, large-scale MIMO radar waveform design remains an open problem.

In this paper, we propose Riemannian manifold optimization algorithms for the beampattern matching problem with good correlation properties in both normal-scale and large-scale MIMO radar scenarios\footnote{\cite{Arxiv} is a preprint uploaded by the author of this manuscript to arxiv.org and has not been published. It is important to note that \cite{Arxiv} only provides a preliminary idea for the large-scale scenario without conducting in-depth research or numerical simulations. The complete work is presented in this manuscript.}, as presented in \cite{L-BFGS} and \cite{consensus-ADMM}. Compared to the algorithms in the aforementioned works, the proposed algorithms do not require constraint relaxing, which improve the efficiency of obtaining the solution. They also converge faster and provide theoretical convergence guarantees. This allows the optimized waveform to better match the desired spatial beampattern, while more effectively suppressing the spatial auto/cross-correlation levels in MIMO radar systems.
The main contributions are as follows:
\begin{itemize}
\item \emph{Manifold and related operator construction}:
We analyze the geometric structure of the unimodular matrix set and construct the corresponding unimodular (UM) manifold, allowing the optimization problem to be solved without vectorizing the matrix-form variable into vector form, thereby reducing the computational complexity.
By embedding constraint into the search space, the original non-convex optimization problem is reformulated and transformed as an unconstrained optimization problem on the manifold.
The constrained space corresponds to a product manifold obtained by the Cartesian product of the UM manifold and the linear manifold associated with Euclidean space. Additionally, we develop the operators required for optimization algorithm design, including Riemannian gradient, retraction, and vector transport.

\item \emph{Low-complexity solving algorithm in the normal-scale case}:
We customize the unimodular manifold gradient descent (UM-GD) algorithm to solve the waveform design problem in the normal-scale case. In the implementation, the gradient computation for each iteration is adjusted by merging summation terms, achieving low-complexity computations. Then, based on UM-GD, its accelerated version of the unimodular manifold accelerated gradient descent (UM-AGD) algorithm is proposed to speed up the convergence.

\item \emph{Stochastic Gradient solving algorithm in the large-scale case}:
We observe that in the large-scale case, the computational cost of the UM-GD and UM-AGD algorithms increases significantly, and the gradient summation merging operation results in a very large data storage requirement. To address this, we reformulate the waveform design problem as a summation of multiple loss functions and customize a unimodular manifold stochastic variance reduced gradient (UM-SVRG) algorithm to solve it.
This significantly reduces the computational complexity of the algorithm in the large-scale case, effectively improving its computational feasibility, and, compared to UM-GD and UM-AGD, removes the need to store large-scale matrix.

\item \emph{Theoretically-guaranteed performance}:
We prove that the proposed UM-GD, UM-AGD, and UM-SVRG algorithms are guaranteed convergent to a stationary point of the non-convex optimization problem if proper parameters are chosen.
\end{itemize}

The rest of the paper is organized as follows: In Section II, we present the optimization problem corresponding to the beampattern design problem and transform it into an unconstrained problem on the manifold for solution. In Section III, we analyze the geometric structure of the constraint space and construct the corresponding manifold, along with the operators required for the customized optimization algorithm. In Section IV, we customize the UM-GD algorithm on the manifold for the normal-scale case and propose an accelerated version, the UM-AGD algorithm. For the large-scale case, we customize the UM-SVRG algorithm for the solution. In Section V, we provide the performance analysis of the proposed UM-GD, UM-AGD and UM-SVRG algorithms, including convergence and computational complexity. Finally, in Section VI, simulation results demonstrate the effectiveness of the proposed algorithms, and Section VII presents the conclusion.

{\it Notations:} scalar variables are represented by standard lowercase letters, vectors and matrices are denoted by lowercase and uppercase bold symbols. $\mathbb{R}$ and $\mathbb{C}$ denote the sets of real and complex values. The superscripts $(\cdot)^*,(\cdot)^T$ and $(\cdot)^H$ denote conjugate operator, transpose operator, and conjugate transpose operator respectively. $\operatorname{Tr}(\cdot)$, $\mathbb{E}(\cdot)$ represent matrix trace, and expectation. $|\cdot|$ denotes the absolute value and $\|\cdot\|_2$ denotes the Euclidean vector norm. $\mathbf{I}_m$ denotes the $m \times m$ identity matrix. $\nabla \left(\cdot\right)$ represents the Euclidean gradient of a function. $\operatorname{grad}\left(\cdot\right)$ represents the Riemannian gradient of a function. $\operatorname{Re}\left(\cdot\right)$ takes the real part of the complex variable. $\operatorname{Im}\left(\cdot\right)$ takes the image part of the complex variable. $\circ$ denotes the Hadamard product and $\times$ denotes the Cartesian product. vec $(\cdot)$ vectorizes a matrix by stacking its columns on top of one another.
$\operatorname{dim}\left(\cdot\right)$ denotes the dimension of the space, and $\operatorname{ker}\left(\cdot\right)$ denotes the kernel of matrix.

\section{System Model and Problem Formulation}

\begin{figure}[htbp]
\centerline{\includegraphics[scale=0.53]{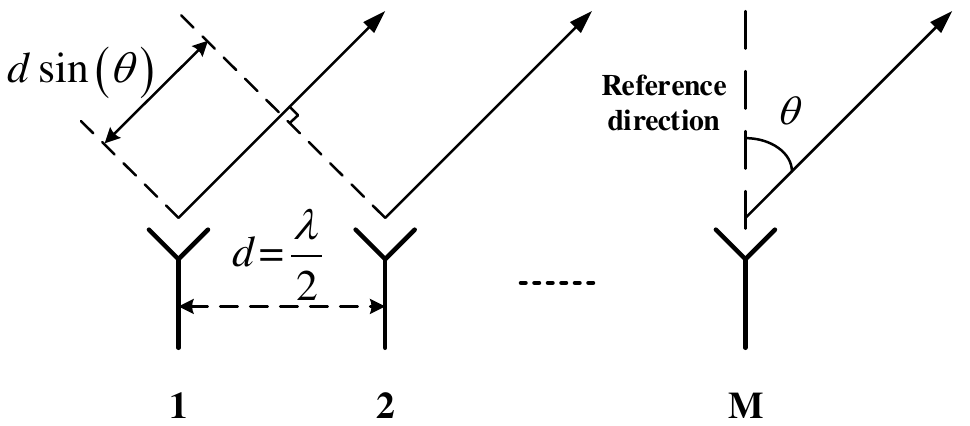}}
\caption{Uniform Linear Array MIMO radar system equipped with $M$ antennas. }
\label{model}
\end{figure}
As shown in Fig. ~\ref{model}, we consider a MIMO radar system equipped with $M$ transmitting elements. The length of the waveform transmitted by each element is $N$ with $N>M$. $\mathbf{X} \in \mathbb{C}^{N \times M}$ is the waveforms transmitted by the MIMO radar system.

The design of the MIMO radar transmission pattern employs the pattern matching criterion, constructing a least squares problem to align the actual transmission pattern with the desired pattern, suppressing clutter components to achieve higher antenna gain. Simultaneously, the objective function minimizes spatial correlation to enhance spatial resolution and suppress clutter components. In practical radar system applications, to maximize transmitter efficiency and avoid nonlinear distortion of the power amplifier, the transmitted waveform is required to maintain unimodular property.

In this work, the design of MIMO radar probing waveform needs to consider the following objectives\cite{consensus-ADMM}.
\begin{itemize}
\item Matching desired beampattern:
\end{itemize}
\begin{equation}\label{Beampattern}
e(\alpha, \mathbf{X})=\sum_{\theta \in \Theta}\left|\alpha \bar{P}_\theta-\mathbf{a}_\theta^H \mathbf{X}^H \mathbf{X} \mathbf{a}_\theta\right|^2,
\end{equation}
where $\theta$ belongs to angle set $\Theta = \left(-90^{\circ}, 90^{\circ}\right)$, $\bar{P}_\theta$ is the desired beampattern, $\mathbf{a}_{\theta}$ is the steering vector at angle $\theta$ used to describe the signal direction characteristics, and $\alpha$ is a scaling variable that needs to be optimized.

\begin{itemize}
\item Minimizing spatial correlation level:
\end{itemize}
\begin{equation}\label{correlation}
P(\mathbf{X})=\sum_{\tau \in \mathcal{D} \backslash 0} \sum_{ \theta_i \in \hat{\Theta}} \left|P_{\theta_i, \theta_i, \tau}\right|^2+\sum_{\tau \in \mathcal{D}} \sum_{\substack{\theta_i \neq \theta_j \\ \theta_i, \theta_j \in \hat{\Theta}}} \left|P_{\theta_i, \theta_j, \tau}\right|^2
\end{equation}
where $\mathcal{D}$ is the time delay parameter set of interest, $\hat{\Theta}$ is the set of angles of interest, and $P_{\theta_i,\theta_j,\tau}$ represents the spatial correlation between the synthesized signal at angle $\theta_i$ and the synthesized signal at angle $\theta_j$ delayed by $\tau$ moments, and $P_{\theta_i,\theta_i,\tau}$ is similarly defined. $P_{\theta_i,\theta_j,\tau}$ is expressed as follow
\begin{equation}
P_{\theta_i,\theta_j,\tau}=\mathbf{a}_{\theta_i}^H\mathbf{X}^H\mathbf{S}_{\tau}\mathbf{X}\mathbf{a}_{\theta_j},
\end{equation}
where $\mathbf{S}_{\tau}\mathbf{X}\mathbf{a}_{\theta}$ is used to describe the synthesized signal at angle $\theta$ delayed by $\tau$ moments. $\mathbf{S}_\tau \in \mathbb{R}^{N\times N}$ is the time delay shift matrix given by
\begin{equation}
\begin{split}
 &\qquad\qquad \quad \tau~{\rm zeros}\\
&{\bf{S}}_{\tau}=
\left[
  \begin{array}{cccccc}
    ~ & \overbrace{0~ \cdots ~0} & 1 & ~ & \mathbf{\scalebox{1.5}0} & ~ \\
    ~ & ~ & ~ &  \ddots &~ & ~  \\
    ~ & ~ & ~ & ~ & 1 & ~ \\
    ~ & \mathbf{\scalebox{2.5}0} & ~ & ~ & ~ & ~  \\
  \end{array}
\right],
\end{split}
\end{equation}
where $\tau$ represents the time delay.

\begin{itemize}
\item Unimodular waveform:
\end{itemize}
\begin{equation}
\left|x_{m}(n)\right|=1, \quad m=1, \cdots, M, n=1, \cdots, N ,
\end{equation}
where $x_{m}(n), n = 1, \cdots, N, m = 1, \cdots, M$ is the element in the $n$-th row and $m$-th column of the matrix $\mathbf{X}$, representing the signal transmitted by the $m$-th transmitting element at the $n$-th moment.

By jointly optimizing the above objectives and constraint, the design of MIMO radar probing waveforms can be formulated as the following optimization problem
\begin{equation}\label{original optimization problem}
\begin{aligned}
\min _{\alpha, \mathbf{X}} & \quad e(\alpha, \mathbf{X})+w_c^2 P(\mathbf{X}), \\
\text { s.t.}& \quad \left|x_{m}(n)\right|=1, m=1, \ldots, M, n=1, \ldots, N,\\
\end{aligned}
\end{equation}
where $w_c$ is a preset positive weight.
By analyzing problem \eqref{original optimization problem}, we find that the unimodular constraint is a non-convex constraint, and the objective function is a quartic polynomial with respect to variable $\mathbf{X}$. Moreover, the matrix $\mathbf{S_{\tau}}$ is not positive semidefinite, making the objective function non-convex. Mathematically, such non-convex quartic programming problem is classified as NP-hard problem and is difficult to solve directly.

In summary, the key to solving optimization problem \eqref{original optimization problem} lies in customizing an efficient and theoretically guaranteed solution approach.

Riemannian manifold optimization based on Riemannian geometry theory can effectively utilize the intrinsic geometric structure of the unimodular constraint, embedding it into the search space, and transforming the original problem into an unconstrained optimization on the manifold. The original optimization problem can be transformed into an unconstrained problem
\begin{equation}\label{Riemannian optimization problem}
\min _{\left(\alpha, \mathbf{X}\right) \in \mathcal{M}} \quad e(\alpha, \mathbf{X})+P(\mathbf{X}).
\end{equation}

It is import to note that by embedding the constraint of problem \eqref{original optimization problem} into the search space to construct the corresponding manifold, problem \eqref{original optimization problem} can be equivalently transformed into the unconstrained optimization problem \eqref{Riemannian optimization problem} on the manifold. Therefore, the solution to problem \eqref{Riemannian optimization problem} must satisfy the constraint of problem \eqref{original optimization problem}. The geometric properties of the manifold $\mathcal{M}$, as well as the operators required for the customized optimization algorithm based on these geometric properties, will be presented in the next section.

\section{Analysis of the Formulated Riemannian Product Manifold}
In this section, we construct the manifold $\mathcal{M}$ as $\mathbb{R} \times \operatorname{UM}\left(N,M\right)$ such that $\left(\alpha, \mathbf{X}\right) \in \mathcal{M}$ and develop the necessary operators for the subsequent customized optimization algorithm: Riemannian gradient, retraction, and vector transport. Specifically, the Riemannian gradient provides the descent direction on the manifold, retraction ensures that each optimization step remains on the manifold, and vector transport enables the addition of directions at different points.
\subsection{The Geometric Structure of Unimodular Constraint}
By \eqref{Riemannian optimization problem}, we observe that the manifold $\mathcal{M}$ is the product manifold obtained by the Cartesian product of $\mathbb{R}$ and $\operatorname{UM}\left(N, M\right)$. In this subsection, we first construct the manifold $\operatorname{UM}\left(N, M\right)$ corresponding to the unimodular constraint in \eqref{original optimization problem}, and then develop the necessary operators.

The unimodular constraint $\left|x_{m}(n)\right|=1, \quad m=1, \cdots, M, n=1, \cdots, N$ is a subset of $\mathbb{C}^{N \times M}$. We first analyze the property of $\mathbb{C}^{N \times M}$, it is easy to get that the set $\mathbb{C}^{N \times M}$ is a vector space. 
According to \cite{manifold-optimization-absil}, $\mathbb{C}^{N \times M}$ is a linear manifold.
Furthermore, the dimension of linear manifold $\mathbb{C}^{N \times M}$ is $2MN$\cite{differential-geometry}.

\begin{proposition}\label{UM-manifold}
The set
\begin{equation}\label{UM-definition}
\begin{aligned}
&\operatorname{UM}\left(N,M\right):= \\
&\left\{\mathbf{X} \! \in \! \mathbb{C}^{N \times M} \!\! :\!\left|x_{m}(n)\right|\!=\!1, m \! = \! 1, \! \cdots \!, M, n \! = \!1, \! \cdots \!, N \right\},
\end{aligned}
\end{equation}
is a embedded submanifold of $\mathbb{C}^{N \times M}$, and $\operatorname{dim}\left(\operatorname{UM}\left(N,M\right)\right)=MN$.
\end{proposition}

{\it Proof:} This proof is presented in Appendix \ref{proof-UM-manifold}.

The set of tangent vectors at the point $\mathbf{X} \in \operatorname{UM}\left(N, M\right)$ is called the tangent space at $\mathbf{X}$, which provides the search direction for optimization on the manifold.
\begin{lemma}\label{lemma-tangent-space}
The tangent space to $\operatorname{UM}\left(N,M\right)$ is given by
\begin{equation}\label{tangent-space}
 T_{\mathbf{X}}\operatorname{UM}\left(N,M\right)\!=\!\left\{\boldsymbol{\xi}_{\mathbf{X}} \in \mathbb{C}^{N \times M}:  \operatorname{Re}\left(\boldsymbol{\xi}_{\mathbf{X}} \circ \mathbf{X}^* \right)=\mathbf{0}_{N \times M}\right\}.
\end{equation}
\end{lemma}

By endowing tangent space with inner product, we construct a Riemannian structure on manifold $\operatorname{UM}\left(N,M\right)$, and it becomes a Riemannian manifold. Since the manifold $\operatorname{UM}\left(N,M\right)$ is an embedded submanifold of $\mathbb{C}^{N \times M}$, its Riemannian metric is induced by that of $\mathbb{C}^{N \times M}$.
\begin{lemma}\label{lemma-Riemannian-metric}
The Riemannian metric on manifold $\operatorname{UM}\left(N,M\right)$ can be expressed by
\begin{equation}\label{Riemannian-metric}
g_{\mathbf{X}}\left(\boldsymbol{\xi}_{\mathbf{X}},\boldsymbol{\eta}_{\mathbf{X}}\right)=\left\langle \boldsymbol{\xi}_{\mathbf{X}}, \boldsymbol{\eta}_\mathbf{X}\right\rangle_{\mathbf{X}}=\operatorname{Re}\left(\operatorname{Tr}
\left(\boldsymbol{\xi}_{\mathbf{X}}^H\boldsymbol{\eta}_{\mathbf{X}}\right)\right),
\end{equation}
where $\boldsymbol{\xi}_{\mathbf{X}}, \boldsymbol{\eta}_{\mathbf{X}} \in T_{\mathbf{X}}\operatorname{UM}\left(N,M\right)$, and $\left\langle \boldsymbol{\xi}_{\mathbf{X}}, \boldsymbol{\eta}_\mathbf{X}\right\rangle_{\mathbf{X}}$ denotes the inner product of two elements $\boldsymbol{\xi}_{\mathbf{X}}$ and $\boldsymbol{\eta}_{\mathbf{X}}$ of $ T_{\mathbf{X}}\operatorname{UM}\left(N,M\right)$.
\end{lemma}

{\it Proof:} The proof of Lemma \ref{lemma-tangent-space} and lemma \ref{lemma-Riemannian-metric} can be found in Appendix \ref{proof-tangent}.

\subsection{Operator Construction on UM manifold}
In the previous subsection, it is mentioned that a tangent vector in $T_{\mathbf{X}}\operatorname{UM}\left(N,M\right)$ can be graphically understood as an ``arrow'' tangent to manifold $\operatorname{UM}\left(N,M\right)$. In this subsection, we present the calculation method for the Riemannian gradient, which represents the direction of steepest ascent for a given objective function, as well as the retraction map from $T_{\mathbf{X}}\operatorname{UM}\left(N,M\right)$ to the manifold $\operatorname{UM}\left(N,M\right)$. This retraction ensures that each iteration stays on the manifold. Since tangent vectors in different tangent spaces belong to different vector spaces and cannot be directly operated upon, they need to be transferred to a common vector space for processing. Finally, we introduce the vector transport operator between two tangent spaces.
\subsubsection{Riemannian Gradient}
Assume a smooth real-valued function $f\left(\mathbf{X}\right)$ is on manifold $\operatorname{UM}\left(N,M\right)$, and its Riemannian gradient $\operatorname{grad}f\left(\mathbf{X}\right)$ at $\mathbf{X}$ is given by
\begin{equation}
\operatorname{grad}f\left(\mathbf{X}\right)=\operatorname{P}_{\mathbf{X}}\nabla f\left(\mathbf{X}\right),
\end{equation}
where $\nabla f\left(\mathbf{X}\right)$ is the Euclidean gradient of $f\left(\mathbf{X}\right)$, and $\operatorname{P}_{\mathbf{X}}\nabla f\left(\mathbf{X}\right)$ denotes the orthogonal projective from $\nabla f\left(\mathbf{X}\right)$ to $T_{\mathbf{X}}\operatorname{UM}\left(N,M\right)$.
\begin{lemma}\label{Riemannian-gradient}
For manifold $\operatorname{UM}\left(N,M\right)$, the Riemannian gradient is given by
\begin{equation}\label{Riemannian-gradient-operator}
\operatorname{grad}f\left(\mathbf{X}\right)=\nabla f\left(\mathbf{X}\right)-\operatorname{Re}\left(\nabla f\left(\mathbf{X}\right)\circ \mathbf{X}^*\right)\circ \mathbf{X}.
\end{equation}
\end{lemma}

{\it Proof:} This proof is presented in Appendix \ref{proof-Riemannian-gradient}.

\subsubsection{Retraction}
A retraction on manifold $\operatorname{UM}\left(N,M\right)$ at $\mathbf{X}$, denotes by $R_{\mathbf{X}}$, is a mapping from $T_{\mathbf{X}}\operatorname{UM}\left(N,M\right)$ to $\operatorname{UM}\left(N,M\right)$ with a local rigidity condition that preserves gradient at $\mathbf{X}$.
\begin{lemma}\label{Retraction}
The retraction on $\operatorname{UM}\left(N,M\right)$ is given by
\begin{equation}\label{retraction-express}
R_{\mathbf{X}}\left(\boldsymbol{\xi}_{\mathbf{X}}\right)=\left(\mathbf{X}+\boldsymbol{\xi}_{\mathbf{X}}\right) \circ \frac{1}{\left|\mathbf{X}+\boldsymbol{\xi}_{\mathbf{X}}\right|_{\left(n,m\right)}},
\end{equation}
where $\mathbf{X} \in \operatorname{UM}\left(N,M\right)$, and $\boldsymbol{\xi}_{\mathbf{X}} \in T_{\mathbf{X}}\operatorname{UM}\left(N,M\right)$.
The specific form of $\frac{1}{\left|\mathbf{X}+\boldsymbol{\xi}_{\mathbf{X}}\right|_{\left(n,m\right)}}$ is as follows
\begin{equation}
\begin{aligned}
&\frac{1}{\left|\mathbf{X}+\boldsymbol{\xi}_{\mathbf{X}}\right|_{\left(n,m\right)}}\\
&=\left[\begin{array}{cccc}
\frac{1}{\left|x_1(1)+\xi_{11}\right|} & \frac{1}{\left|x_2(1)+\xi_{12}\right|} & \cdots & \frac{1}{\left|x_M(1)+\xi_{1M}\right|} \\
\frac{1}{\left|x_1(2)+\xi_{21}\right|} & \frac{1}{\left|x_2(2)+\xi_{22}\right|} & \cdots & \frac{1}{\left|x_M(2)+\xi_{2M}\right|} \\
\vdots & \vdots & \ddots & \vdots \\
\frac{1}{\left|x_1(N)+\xi_{N1}\right|} & \frac{1}{\left|x_2(N)+\xi_{N2}\right|} & \cdots & \frac{1}{\left|x_M(N)+\xi_{NM}\right|}
\end{array}\right].
\end{aligned}
\end{equation}
\end{lemma}

{\it Proof:} This proof is presented in Appendix \ref{proof-Retraction}.
\subsubsection{Vector Transport}
The vector transport $\mathcal{T}$ on manifold $\operatorname{UM}\left(N,M\right)$ specifies how to transport a tangent vector $\xi_{\mathbf{X}}$ from $T_{\mathbf{X}}\operatorname{UM}\left(N,M\right)$ to $T_{R_{\mathbf{X}}\left(\boldsymbol{\eta}_{\mathbf{X}}\right)}\operatorname{UM}\left(N,M\right)$. We define the vector transport as
\begin{equation}\label{vector-transport-projection}
\mathcal{T}_{\boldsymbol{\eta}_{\mathbf{X}}} \boldsymbol{\xi}_{\mathbf{X}}=\mathrm{P}_{R_{\mathbf{X}}\left(\boldsymbol{\eta}_{\mathbf{X}}\right)} \boldsymbol{\xi}_{\mathbf{X}}.
\end{equation}

\begin{lemma}
According to Lemma \ref{Riemannian-gradient}, the vector transport on $\operatorname{UM}\left(N,M\right)$ is defined by
\begin{equation}
\mathcal{T}_{\boldsymbol{\eta}_{\mathbf{X}}}\boldsymbol{\xi}_{\mathbf{X}}=\boldsymbol{\xi}_{\mathbf{X}}-\operatorname{Re}\left(\boldsymbol{\xi}_{\mathbf{X}} \circ \left(R_{\mathbf{X}}\left(\boldsymbol{\eta}_{\mathbf{X}}\right)\right)^*\right) \circ R_{\mathbf{X}}\left(\boldsymbol{\eta}_{\mathbf{X}}\right).
\end{equation}
\end{lemma}

\subsection{Operator Construction on manifold $\mathcal{M}$}   
In this subsection, based on the optimization variables $\left(\alpha, \mathbf{X}\right) \in \mathcal{M}$ in optimization problem \eqref{Riemannian optimization problem}, we construct the manifold $\mathcal{M}$ as the product manifold obtained from the Cartesian product of the manifolds $\mathbb{R}$ and $\operatorname{UM}\left(N, M\right)$, denotes as $\mathbb{R} \times \operatorname{UM}\left(N, M\right)$. Then, based on the operators on the manifold $\operatorname{UM}\left(N, M\right)$ constructed in the previous subsection, we construct the operators on $\mathbb{R} \times \operatorname{UM}\left(N, M\right)$.

The set $\mathbb{R} \times \operatorname{UM}\left(N,M\right)$ is defined as the set of pairs $\left(\alpha, \mathbf{X}\right)$, where $\alpha$ is in $\mathbb{R}$ and $\mathbf{X}$ is in $\operatorname{UM}\left(N,M\right)$. Then, $\mathbb{R} \times \operatorname{UM}\left(N,M\right)$ is a manifold of dimension $MN+1$ with tangent spaces given by
\begin{equation}
T_{\left(\alpha, \mathbf{X}\right)}\left(\mathbb{R} \times \operatorname{UM}\left(N,M\right)\right)=T_{\alpha}\mathbb{R} \times T_{\mathbf{X}}\operatorname{UM}\left(N,M\right).
\end{equation}

\subsubsection{Riemannian Gradient}
We let $f\left(\alpha, \mathbf{X}\right) = e(\alpha, \mathbf{X})+P_c(\mathbf{X})$. For smooth function $f: \mathbb{R} \times \operatorname{UM}\left(N,M\right) \rightarrow \mathbb{R}: \left(\alpha, \mathbf{X}\right) \rightarrow f\left(\alpha, \mathbf{X}\right)$, the Riemannian gradient is given by
\begin{equation}\label{product-Riemannian-gradient}
\operatorname{grad}f\left(\alpha, \mathbf{X}\right)=\left(\nabla_{\alpha} f\left(\alpha, \mathbf{X}\right), \operatorname{grad}_{\mathbf{X}}f\left(\alpha, \mathbf{X}\right)\right),
\end{equation}
where $\nabla_{\alpha} f\left(\alpha, \mathbf{X}\right)$ is the Euclidean gradient of $f\left(\alpha, \mathbf{X}\right)$ with respect to $\alpha$ when $\mathbf{X}$ is fixed, $\operatorname{grad}_{\mathbf{X}}f\left(\alpha, \mathbf{X}\right)=\operatorname{P}_{\mathbf{X}}\nabla_{\mathbf{X}}f\left(\alpha, \mathbf{X}\right)$, and $\nabla_{\mathbf{X}}f\left(\alpha, \mathbf{X}\right)$ is the Euclidean gradient of $f\left(\alpha, \mathbf{X}\right)$ with respect to $\mathbf{X}$ when $\alpha$ is fixed. The specific expression for $\operatorname{grad}_{\mathbf{X}}f\left(\alpha, \mathbf{X}\right)$ can be found in \eqref{Riemannian-gradient-operator}.

\subsubsection{Retraction}
$\operatorname{UM}\left(N,M\right)$, $\mathbb{R}$ are equipped with retractions $R_{\mathbf{X}}, R_{\alpha}$. The map $R_{\alpha, \mathbf{X}}: T_{\left(\alpha, \mathbf{X}\right)}\left(\mathbb{R} \times \operatorname{UM}\left(N, M\right)\right) \rightarrow   \mathbb{R} \times \operatorname{UM}\left(N, M\right) $ is the retraction on manifold  $ \mathbb{R} \times \operatorname{UM}\left(N, M\right)$ \cite{manifold-nicolas}, which is given by
\begin{equation}
\begin{aligned}
R_{\left(\alpha, \mathbf{X}\right)}\left(\xi_{\alpha}, \boldsymbol{\xi}_{\mathbf{X}}\right)&=\left(R_{\alpha}\left(\xi_{\alpha}\right), R_{\mathbf{X}}\left(\boldsymbol{\xi}_{\mathbf{X}}\right)\right)\\
&=\left(\alpha+\xi_{\alpha}, R_{\mathbf{X}}\left(\boldsymbol{\xi}_{\mathbf{X}}\right)\right).
\end{aligned}
\end{equation}

\subsubsection{Vector Transport}
According to \eqref{vector-transport-projection} and \eqref{product-Riemannian-gradient}, we can obtain
\begin{equation}\label{vector-transport-M}
\begin{aligned}
\mathcal{T}_{\left(\eta_{\alpha}, \boldsymbol{\eta}_{\mathbf{X}}\right)} \left(\xi_{\alpha}, \boldsymbol{\xi}_{\mathbf{X}}\right)&=\mathrm{P}_{R_{\left(\alpha, \mathbf{X}\right)}\left(\eta_{\alpha}, \boldsymbol{\eta}_{\mathbf{X}}\right)} \left(\xi_{\alpha}, \boldsymbol{\xi}_{\mathbf{X}}\right)\\
&=\left(\xi_{\alpha}, \mathcal{T}_{\boldsymbol{\eta}_{\mathbf{X}}}\boldsymbol{\xi}_{\mathbf{X}}\right).
\end{aligned}
\end{equation}

Here, $\xi_\alpha$ remains unchanged after vector transport, and later we simplify $\mathcal{T}_{\left(\eta_{\alpha}, \boldsymbol{\eta}_{\mathbf{X}}\right)} \left(\xi_{\alpha}, \boldsymbol{\xi}_{\mathbf{X}}\right)$ to $\mathcal{T}_{ \boldsymbol{\eta}_{\mathbf{X}}} \left(\xi_{\alpha}, \boldsymbol{\xi}_{\mathbf{X}}\right)$.

\begin{remark}
In the customized optimization algorithm in the next section, we obtain the optimization descent direction for problem $\eqref{Riemannian optimization problem}$ using the Riemannian gradient on the manifold $\operatorname{UM}\left(N, M\right)$. Retraction ensures that each optimization step remains on the manifold $\mathcal{M}$, and vector transport enables the addition of descent directions at different points.
\end{remark}

\section{Proposed Algorithm}
 By observing \eqref{Riemannian optimization problem}, we can see that both $e\left(\alpha, \mathbf{X}\right)$ and $P\left(\mathbf{X}\right)$ are in the form of sums of multiple quartic terms. Directly solving the unconstrained problem on this manifold without any preprocessing requires repeatedly calculating and summing gradients at each iteration. This approach is acceptable when the number of summands is relatively small. When the number of summands is large, the computational cost per iteration becomes very high.
 Therefore, in this section, we consider both normal-scale and large-scale (typically $M>2^6$) cases and design specific optimization strategies for each case to reduce the computational cost. Here, normal-scale and large-scale can be simply understood as MIMO and massive MIMO, respectively.

\subsection{Case I: Normal-Scale Beampattern Problem}
In the normal-scale case, the value of $M$, $N$, and $\tau$ are relatively small. At this point, the influence of the summation terms in $e\left(\alpha, \mathbf{X}\right)$ on the computational cost of each iteration is significant. We use a full gradient algorithm to solve the problem and combine the multiple summation terms in $\nabla e\left(\alpha, \mathbf{X}\right)$ for each iteration through matrix operations.

\subsubsection{UM-GD Algorithm}
The steepest descent method on manifold $\mathcal{M}$ is an extension of the one in Euclidean space. In the specific algorithm design, we aim to merge the summation terms as much as possible. Let $f\left(\alpha, \mathbf{X}\right) = e(\alpha, \mathbf{X})+P(\mathbf{X})$, we first analyze the Euclidean gradient of $e(\alpha, \mathbf{X})$ and $P(\mathbf{X})$ as follows
\begin{subequations}
\begin{align}
\hspace{-0.2cm}&\nabla\!_\alpha e(\alpha,\! \mathbf{X})\!=\!\!2\!\left(\!\alpha\! \sum_{\theta \in \Theta}\left|\bar{P}_\theta\right|^2\!\!-\!\!\operatorname{Tr}\!\left(\!\mathbf{X}^H \mathbf{X}\! \sum_{\theta \in \Theta} \bar{P}_\theta \mathbf{a}_\theta \mathbf{a}_\theta^H\!\!\right)\!\!\right) ,\label{Euclidean-gradient-alpha} \\
&\nabla_{\mathbf{X}} e(\alpha, \mathbf{X})=-2 \alpha \mathbf{X} \sum_{\theta \in \Theta} \bar{P}_\theta \notag \mathbf{a}_\theta \mathbf{a}_\theta^H+ \nonumber \\
&2 \operatorname{unvec}\left(\left(\mathbf{I}_M \otimes \mathbf{X}\right)\left(\sum_{\theta \in \Theta} \mathbf{A}_\theta \mathbf{A}_\theta{ }^H\right) \operatorname{vec}\left(\mathbf{X}^H \mathbf{X}\right)\right) ,\label{Euclidean-gradient-f1-X} \\
&\nabla_{\mathbf{X}} P(\mathbf{X})=\sum_{\tau \in \mathcal{D} \backslash 0} \sum_{\theta_i \in \hat{\Theta}}  \frac{\partial\left|P_{\theta_i, \theta_i, \tau}\right|^2} {\partial\mathbf{X}^*}+
\nonumber \nonumber \\
&\hspace{2.0cm}\sum_{\tau \in \mathcal{D} } \sum_{\substack{\theta_i, \theta_j \\ \theta_i, \theta_j \in \hat{\Theta}}}  \frac{\partial\left|P_{\theta_i, \theta_j, \tau}\right|^2}{\partial \mathbf{X}^*}, \label{Euclidean-gradient-f2-X}
\end{align}
\end{subequations}
where $\mathbf{A}_{\theta} = \operatorname{vec}\left(\mathbf{a}_{\theta}\mathbf{a}_{\theta}^H\right)$ and
\begin{subequations}
\begin{align}
\frac{\partial\left|P_{\theta_i, \theta_i, \tau}\right|^2}{\partial \mathbf{X}^*} &=  \notag \left(\operatorname{Tr}\left(\mathbf{S}_{\tau}\mathbf{X}\mathbf{a}_{\theta_i}\mathbf{a}_{\theta_i}^H\mathbf{X}^H\right)\right)^*\mathbf{S}_{\tau}\mathbf{X}\mathbf{a}_{\theta_i}\mathbf{a}_{\theta_i}^H\\
&+\operatorname{Tr}\left(\mathbf{S}_{\tau}\mathbf{X}\mathbf{a}_{\theta_i}\mathbf{a}_{\theta_i}^H\mathbf{X}^H\right)\mathbf{S}_{\tau}^T\mathbf{X}\mathbf{a}_{\theta_i}\mathbf{a}_{\theta_i}^H, \label{p-norm2-gradient-specific1} \\
\frac{\partial\left|P_{\theta_i, \theta_j, \tau}\right|^2}{\partial \mathbf{X}^*} &=  \notag \left(\operatorname{Tr}\left(\mathbf{S}_{\tau}\mathbf{X}\mathbf{a}_{\theta_j}\mathbf{a}_{\theta_i}^H\mathbf{X}^H\right)\right)^*\mathbf{S}_{\tau}\mathbf{X}\mathbf{a}_{\theta_j}\mathbf{a}_{\theta_i}^H\\
&+\operatorname{Tr}\left(\mathbf{S}_{\tau}\mathbf{X}\mathbf{a}_{\theta_j}\mathbf{a}_{\theta_i}^H\mathbf{X}^H\right)\mathbf{S}_{\tau}^T\mathbf{X}\mathbf{a}_{\theta_j}\mathbf{a}_{\theta_i}^H. \label{p-norm2-gradient-specific2}
\end{align}
\end{subequations}
{\it Proof:} This proof is presented in Appendix \ref{proof-euclidean-gradient}.

We can obtain the update rule for UM-GD as follows
\begin{equation}\label{R-GD-update}
\left(\alpha_{k+1}, \mathbf{X}_{k+1}\right) = R_{\left(\alpha_k, \mathbf{X}_k\right)}\left(-t_k \operatorname{grad}f\left(\alpha_k, \mathbf{X}_k\right)\right),
\end{equation}
where $t_k \in \mathbb{R}$ is the step size, and the specific form of $\operatorname{grad}f\left(\alpha_k, \mathbf{X}_k\right)$ is obtained by substituting \eqref{Euclidean-gradient-alpha}, \eqref{Euclidean-gradient-f1-X}, and \eqref{Euclidean-gradient-f2-X} into \eqref{product-Riemannian-gradient}.

\subsubsection{Acceleration Strategy}
In the proposed UM-GD algorithm, the choice of the step size is crucial.
Too small a step will result in slow convergence, while too large a step will make the optimization algorithm difficult to converge. A popular condition states that the step size $t_k$ should first provide a sufficient decrease in the objective function $ f(\alpha, \mathbf{X})$.

Given scalars $\bar{t}>0, \beta, \sigma \in \left(0,1\right)$, we choose {$t_k = \beta^{\bar{m}}\bar{t}$, where $\bar{m}$ is the smallest nonnegative integer such that 
\begin{equation}\label{Accelerate-R-SD-stepsize-condition}
\begin{aligned}
&f\left(\alpha_k, \mathbf{X}_k\right)-f\left(R_{\left(\alpha_k, \mathbf{X}_k\right)}\left(-t_k \operatorname{grad}f\left(\alpha_k, \mathbf{X}_k\right)\right)\right)\geq \\
&-\sigma\left\langle \operatorname{grad}f\left(\alpha_k, \mathbf{X}_k\right), -t_k \operatorname{grad}f\left(\alpha_k, \mathbf{X}_k\right)\right\rangle_{\left(\alpha_k, \mathbf{X}_k\right)}.
\end{aligned}
\end{equation}

In summary, the UM-AGD algorithm for optimization problem \eqref{Riemannian optimization problem} is provided in Table.~\ref{proposed accelerate R-SD algorithm}.
\begin{table}[t]
\renewcommand \arraystretch{1.2}
\caption{The proposed UM-AGD algorithm }
\label{proposed accelerate R-SD algorithm}
\centering
\begin{tabular}{l}
 \hline\hline
 {\bf Parameters:} Scalars $\bar{t}>0$, $\beta, \sigma \in \left(0, 1\right)$. \\
 {\bf Initialize:} $\left(\alpha_0,\mathbf{X}_0\right)\in \mathcal{M}, t_k = \bar{t} \left(k = 1,2,...\right)$.\\
 {\bf Iterate:} for $k = 1,2,...$ \\
  \hspace{0.2cm} Calculate the Riemannian gradient $\operatorname{grad}f\left(\alpha_k, \mathbf{X}_k\right)$ \\
  via \eqref{product-Riemannian-gradient}, \eqref{Euclidean-gradient-alpha}--\eqref{Euclidean-gradient-f2-X}. \\
  \hspace{0.2cm}{\bf while} $f\left(\alpha_k, \mathbf{X}_k\right)-f\left(R_{\left(\alpha_k, \mathbf{X}_k\right)}\left(-t_k \operatorname{grad}f\left(\alpha_k, \mathbf{X}_k\right)\right)\right)$\\
  \hspace{0.5cm}$<-\sigma\left\langle \operatorname{grad}f\left(\alpha_k, \mathbf{X}_k\right), -t_k \operatorname{grad}f\left(\alpha_k, \mathbf{X}_k\right)\right\rangle_{\left(\alpha_k, \mathbf{X}_k\right)}$\\
  \hspace{3cm} $t_k = \beta t_k$.\\
  \hspace{0.2cm}{\bf end while}\\
  \hspace{0.2cm} Update $\left(\alpha_{k+1}, \mathbf{X}_{k+1}\right)$ from $\left(\alpha_{k}, \mathbf{X}_{k}\right)$ as \eqref{R-GD-update}.\\
  {\bf Until} some preset termination criterion is satisfied.\\
  {\bf end} \\
 \hline\hline
\end{tabular}
\end{table}

\subsection{Case II: Large-Scale Beampattern Problem}
In the large-scale case, the number of antennas $M$ is large, while the waveform length $N$ and the number of interested delays $\tau$ are also substantial. The summation terms in $e(\alpha, \mathbf{X})$ and $P(\mathbf{X})$ significantly impact the computational burden of each iteration of the optimization algorithm. Furthermore, as the matrix dimensions increase, the summation merging strategy mentioned for $e(\alpha, \mathbf{X})$ in the normal-scale case requires storing a matrix with dimensions $M^2 \times M^2$, which leads to explosive growth in storage requirements. Therefore, the high cost and imposing storage burden make the solution strategy for the normal-scale case unsuitable for the large-scale case. The key to solving the large-scale beampattern problem is to address the excessive computational burden caused by multiple summation terms without merging them.

By observing \eqref{Beampattern} and \eqref{correlation}, the optimization problem \eqref{Riemannian optimization problem} can be transformed into
\begin{equation}\label{loss-function}
\min _{\left(\alpha, \mathbf{X}\right) \in \mathcal{M}} \quad \sum_{\theta \in \Theta}e_{\theta}(\alpha, \mathbf{X})+\sum_{\tau \in \mathcal{D}}P_{\tau}\left(\mathbf{X}\right),
\end{equation}
where $e_{\theta}(\alpha, \mathbf{X}) = |\alpha\bar{P}_{\theta}-\mathbf{a}_{\theta}^H\mathbf{X}^H\mathbf{X}\mathbf{a}_{\theta}|^2$, $P_{\tau}\left(\mathbf{X}\right) =  \sum_{ \theta_i \in \hat{\Theta}} \left|P_{\theta_i, \theta_i, \tau}\right|^2+ \sum_{\substack{\theta_i \neq \theta_j \\ \theta_i, \theta_j \in \hat{\Theta}}} \left|P_{\theta_i, \theta_j, \tau}\right|^2 $.

To our knowledge, stochastic variance reduction gradient (SVRG) algorithm \cite{E-SVRG} have a significant advantage in minimizing the average of a large but finite number of loss functions, such as \eqref{loss-function}. Here, we customize the UM-SVRG algorithm based on the idea of the R-SVRG\cite{R-SVRG} algorithm to tackle this issue.

The UM-SVRG algorithm is an extension of the aforementioned Euclidean SVRG algorithm into $\mathcal{M}$ manifold. It employs a two-layer loop, where the superscript $``l"$ denotes the current iteration index of the outer loop, the subscript $``q"$ represents the current iteration index of the inner loop within the current outer loop, and there are $m_l$ inner iterations in each outer loop. The algorithm retains $\left(\widetilde{\alpha}^{l-1}, \widetilde{\mathbf{X}}^{l-1}\right) \in \mathcal{M}$ after $l-1$ outer loops to modify the stochastic gradient $\operatorname{grad}f_{i_q^l}\left(\alpha_q^l, \mathbf{X}_q^l\right)$. The modified Riemannian stochastic gradient of $q$th inner iteration of the $l$th outer loop is
\begin{equation}\label{Riemannian-stochastic-gradient}
\begin{aligned}
\boldsymbol{\varsigma}_q^l =& \operatorname{grad}f_{i_q^l}\left(\alpha_q^l, \mathbf{X}_q^l\right)-\mathcal{T}_{\widetilde{\eta}_{\mathbf{X}_q^l}}\left(\operatorname{grad}f_{i_q^l}\left(\widetilde{\alpha}^{l-1}, \widetilde{\mathbf{X}}^{l-1}\right)\right.\\
&-\left.\operatorname{grad}f\left(\widetilde{\alpha}^{l-1}, \widetilde{\mathbf{X}}^{l-1}\right)\right),
\end{aligned}
\end{equation}
where $i_q^l$ is the random gradient subscripts for objective function $f\left(\alpha, \mathbf{X}\right)$ and $\left(\widetilde{\eta}_{\alpha_q^l}, \widetilde{\eta}_{\mathbf{X}_q^l}\right)$ satisfies $R_{\left(\widetilde{\alpha}^{l-1}, \widetilde{\mathbf{X}}^{l-1}\right)}\left(\widetilde{\eta}_{\alpha_q^l}, \widetilde{\eta}_{\mathbf{X}_q^l}\right) = \left(\alpha_q^l, \mathbf{X}_q^l\right)$.

Specifically
\begin{subequations}
\begin{align}
&\nabla_{\alpha} f_{i_q^l}\left(\alpha, \mathbf{X}\right) \!\!=\!\! 2\left(\!\alpha \left|\bar{P}_{\theta_q^l}\right|^2\!-\!\operatorname{Tr}\left(\mathbf{X}^H \mathbf{X} \bar{P}_{\theta_q^l} \mathbf{a}_{\theta_q^l} \mathbf{a}_{\theta_q^l}^H\!\right) \right),\label{stochastic-gradient-alpha} \\
&\nabla_{\mathbf{X}} f_{i_q^l}\left(\alpha, \mathbf{X}\right) = \nabla_{\mathbf{X}} e_{\theta_q^l}(\alpha, \mathbf{X}) + \nabla_{\mathbf{X}} P_{\tau_q^l}(\mathbf{X}) , \label{stochastic-gradient} \\
&\nabla_{\mathbf{X}} e_{\theta_q^l}(\alpha, \mathbf{X})=\!2\!\left(\mathbf{a}_{\theta_q^l}^H\mathbf{X}^H\mathbf{X}\mathbf{a}_{\theta_q^l}\mathbf{X}\mathbf{a}_{\theta_q^l}\mathbf{a}_{\theta_q^l}^H- \alpha \mathbf{X} \bar{P}_{\theta_q^l} \mathbf{a}_{\theta_q^l} \mathbf{a}_{\theta_q^l}^H\right) ,  \label{stochastic-gradient-f1-X} \\
&\nabla_{\mathbf{X}} P_{\tau_q^l}(\mathbf{X})\!\!= \!\!\sum_{\theta_i \in \hat{\Theta}}\!\!  \frac{\partial\left|P_{\theta_i, \theta_i, \tau_q^l}\right|^2}{\partial \mathbf{X}^*}
\!+\!\!\!\sum_{\substack{\theta_i \neq \theta_j \\ \theta_i, \theta_j \in \hat{\Theta}}}\!\! \frac{\partial\left|P_{\theta_i, \theta_j, \tau_q^l}\right|^2}{\partial \mathbf{X}^*} \label{stochastic-gradient-f2-X},
\end{align}
\end{subequations}
where the random gradient subscripts $\theta_q^l$ and $\tau_q^l$ are obtained by randomly selecting from the sets $\Theta$ and $\mathcal{D}$, respectively.

Moreover, according to the Riemannian gradient \eqref{Riemannian-gradient-operator}, \eqref{product-Riemannian-gradient} and the vector transport operator \eqref{vector-transport-M}, we can derive the specific expression of \eqref{Riemannian-stochastic-gradient}.

The update rule for UM-SVRG is as follows
\begin{equation}\label{R-SVRG-update-rule}
\left(\alpha_q^l, \mathbf{X}_q^l\right)=R_{\left(\alpha_{q-1}^l, \mathbf{X}_{q-1}^l\right)}\left(-t_{q-1}^l\boldsymbol{\varsigma}_q^l\right).
\end{equation}

In summary, the UM-SVRG algorithm for optimization problem \eqref{Riemannian optimization problem} is provided in Table.~\ref{proposed R-SVRG algorithm}.
\begin{table}[t]
\renewcommand \arraystretch{1.2}
\caption{The proposed UM-SVRG algorithm }
\label{proposed R-SVRG algorithm}
\centering
\begin{tabular}{l}
 \hline\hline
 {\bf Parameters:} Update frequency $m_l > 0$ and sequence $\{t_q^l\}$ of \\ positive step sizes. \\
 {\bf Initialize:} $\left(\widetilde{\alpha}^0, \widetilde{\mathbf{X}}^0\right)\in \mathcal{M}$ \\
 {\bf Iterate:} for $l = 1,2,...$ \\
  \hspace{0.2cm} Calculate the full Riemannian gradient $\operatorname{grad}f\left( \widetilde{\alpha}^{l-1}, \widetilde{\mathbf{X}}^{l-1}\right)$ via \eqref{product-Riemannian-gradient}. \\
  \hspace{0.2cm} Store $\left(\alpha_0^l, \mathbf{X}_0^l\right) =\left(\widetilde{\alpha}^{l-1}, \widetilde{\mathbf{X}}^{l-1}\right)$. \\
  \hspace{0.2cm} {\bf Iterate:} for $q = 1,2,...,m_l$ \\
  \hspace{0.4cm} Choose $\theta_q^l \in \Theta $ and $\tau_q^l \in \mathcal{D}$ uniformly at random.\\
  \hspace{0.4cm} Calculate the modified Riemannian stochastic gradient $\boldsymbol{\varsigma}_q^l$ as \eqref{Riemannian-stochastic-gradient}\\
  \hspace{0.4cm} via \eqref{stochastic-gradient-alpha}--\eqref{stochastic-gradient-f2-X}.\\
  \hspace{0.4cm} Update $\left(\alpha_1^l, \mathbf{X}_q^l\right)$ from $\left(\alpha_{q-1}^l, \mathbf{X}_{q-1}^l\right)$ as \eqref{R-SVRG-update-rule}.\\
  \hspace{0.2cm} {\bf end}\\
  \hspace{0.2cm} set $\left(\widetilde{\alpha}^l, \widetilde{\mathbf{X}}^l \right)= \left( \alpha_{m_l}^l, \mathbf{X}_{m_l}^l\right)$.\\
  {\bf Until} some preset termination criterion is satisfied.\\
  {\bf end} \\
 \hline\hline
\end{tabular}
\end{table}

\section{Performance Analysis}
In this section, we first perform analyses of the proposed algorithms to provide theoretical guarantees for their convergence. Then, we analyze how the algorithms achieve low complexity and present the computational complexity results.
\subsection{Convergence Analysis}
We have the following theorems to demonstrate the convergence properties of the proposed UM-GD algorithm, UM-AGD algorithm, and UM-SVRG algorithm.

\begin{theorem}\label{R-SD-convergence}
Let $f\left(R_{\left(\alpha, \mathbf{X}\right)}\left(\mathbf{S}\right)\right)$ satisfy
\begin{equation}\label{lipschitz-continuous}
f\left(R_{\left(\alpha, \mathbf{X}\right)}\left(\mathbf{S}\right)\right)\leq f\left(\alpha, \mathbf{X}\right)+\langle\operatorname{grad} f(\alpha, \mathbf{X}), \mathbf{S}\rangle+\frac{L}{2}\|\mathbf{S}\|_{\left(\alpha, \mathbf{X}\right)}^2,
\end{equation}
where $L>0$ is a constant, and $\|\mathbf{S}\|_{\left(\alpha, \mathbf{X}\right)}^2 = g_{\left(\alpha, \mathbf{X}\right)}\left(\mathbf{S}, \mathbf{S}\right)$. If $\left(\alpha_0, \mathbf{X}_0\right), \left(\alpha_1, \mathbf{X}_1\right),..., \left(\alpha_k, \mathbf{X}_k\right)$ generated by UM-GD algorithm with step-size
\begin{equation}
t_k \in (0,2/L),
\end{equation}
then the algorithm produces sufficient decrease. The proposed UM-GD algorithm is convergent and
\begin{equation}\label{station-point}
\lim _{k \rightarrow \infty}\left\|\operatorname{grad} f\left(\alpha_k, \mathbf{X}_k\right)\right\|_{\left(\alpha_k, \mathbf{X}_k\right)}^2=0.
\end{equation}
\end{theorem}

{\it Proof:} This proof is presented in Appendix \ref{proof-R-SD-convergence}.

\begin{remark}
The proposed accelerated UM-GD algorithm produces a sufficient decrease in the objective function, thus the algorithm converges and satisfies \eqref{station-point}.
\end{remark}
\begin{theorem}\label{R-SVRG-convergence}
If the sequence $\{t_q^l\}$ of positive step sizes satisfies
\begin{subequations}
\begin{align}
&\sum_{l=1}^{\infty}\sum_{q=0}^{m_l}\left(t_q^l\right)^2 < \infty, \label{stepsize condition 1} \\
&\sum_{l=1}^{\infty}\sum_{q=0}^{m_l}t_q^l = \infty. \label{stepsize condition 2}
\end{align}
\end{subequations}
Then, $\{f\left(\alpha_q^l, \mathbf{X}_q^l\right)\}$ converges a.s. and $\operatorname{grad}f\left(\alpha_q^l, \mathbf{X}_q^l\right) \rightarrow 0$ a.s. The proposed UM-SVRG algorithm is convergent.
\end{theorem}

{\it Proof:} This proof is presented in Appendix \ref{proof-R-SVRG-convergence}.

\begin{remark}
We choose $t_q^l=\alpha_0\left(1+t_0 \lambda\left\lfloor k_1 / m_l\right\rfloor\right)^{-1}$, where  $t_0, \lambda>0$, $t_0$ is the initial step size, $k_1$ is the total number of iterations depending on $l$ and $q$, and $\lfloor\cdot\rfloor$ denotes the floor function.
\end{remark}

\subsection{Implementation Analysis}
By observing the proposed UM-GD and UM-SVRG algorithms, we can see that the main computational burden is dominated by the calculation of the Euclidean gradient. In the following, we provide the specific computational complexity of the proposed algorithms and analyze how the proposed algorithms achieve low complexity.
\subsubsection{UM-GD}
In the UM-GD algorithm for the normal-scale case, the Euclidean gradients that need to be computed in each iteration include $\nabla_\alpha e(\alpha, \mathbf{X})$, $\nabla_{\mathbf{X}} e(\alpha, \mathbf{X})$, and $\nabla_{\mathbf{X}} P(\mathbf{X})$. Moreover, $\sum_{\theta \in \Theta}\bar{P}_{\theta}\mathbf{a}_{\theta}\mathbf{a}_{\theta}^H$, $\sum_{\theta \in \Theta} \mathbf{A}_\theta \mathbf{A}_\theta^H$ can be precomputed before the iteration.

For $\nabla_\alpha e(\alpha, \mathbf{X})$, obtaining $\mathbf{X}^H\mathbf{X}$ requires no more than $M^2 N$ complex multiplication operations, and obtaining $\operatorname{Tr}\left(\mathbf{X}^H\mathbf{X}\sum_{\theta \in \Theta}\bar{P}_{\theta}\mathbf{a}_{\theta}\mathbf{a}_{\theta}^H\right)$ requires no more than $M^3$ complex multiplication operations.

For $\nabla_{\mathbf{X}} e(\alpha, \mathbf{X})$, obtaining $\mathbf{X}\sum_{\theta \in \Theta}\bar{P}_{\theta}\mathbf{a}_{\theta}\mathbf{a}_{\theta}^H$ requires no more than $M^2N$ complex multiplication operations. Then, since matrix $\mathbf{I}_M \otimes \mathbf{X}$ has a sparse structure, the second term in \eqref{Euclidean-gradient-f1-X} requires no more than $M^4+M^2N$ complex multiplications.

For $\nabla_{\mathbf{X}} P(\mathbf{X})$, we first analyze the computational complexity of $\nabla_{\mathbf{X}}P_{\tau}\left(\mathbf{X}\right)$, where $\nabla_{\mathbf{X}}P_{\tau}\left(\mathbf{X}\right)$ is defined as follow
\begin{equation}
\nabla_{\mathbf{X}} P_{\tau}(\mathbf{X})\!=\!\sum_{\theta_i \in \hat{\Theta}}  \frac{\partial\left|P_{\theta_i, \theta_i, \tau}\right|^2}{\partial \mathbf{X}^*}\!+\!\!\!\sum_{\substack{\theta_i, \theta_j \\ \theta_i \neq \theta_j \in \hat{\Theta}}}   \frac{\partial\left|P_{\theta_i, \theta_j, \tau}\right|^2}{\partial \mathbf{X}^*}. \label{Euclidean-gradient-f2-X-analysis}
\end{equation}

For the calculations of $\operatorname{Tr}\left(\mathbf{S}_{\tau}\mathbf{X}\mathbf{a}_{\theta_i}\mathbf{a}_{\theta_i}^H\mathbf{X}^H\right)$ and $\operatorname{Tr}\left(\mathbf{S}_{\tau}\mathbf{X}\mathbf{a}_{\theta_j}\mathbf{a}_{\theta_i}^H\mathbf{X}^H\right)$, we have
\begin{equation}
\operatorname{Tr}\left(\mathbf{S}_{\tau}\mathbf{X}\mathbf{a}_{\theta_i}\mathbf{a}_{\theta_i}^H\mathbf{X}^H\right) =\mathbf{b}_{\theta_i}^H\mathbf{S}_{\tau}\mathbf{b}_{\theta_i},
\end{equation}
where $\mathbf{b}_{\theta_i} = \mathbf{X}\mathbf{a}_{\theta_i}$ and $\mathbf{b}_{\theta_j} = \mathbf{X}\mathbf{a}_{\theta_j}$.
Since matrix $\mathbf{S}_{\tau}$ exhibits a sparse characteristic. Therefore, the overall computational complexity for all required $\operatorname{Tr}\left(\mathbf{S}_{\tau}\mathbf{X}\mathbf{a}_{\theta_i}\mathbf{a}_{\theta_i}^H\mathbf{X}^H\right)$ and $\operatorname{Tr}\left(\mathbf{S}_{\tau}\mathbf{X}\mathbf{a}_{\theta_j}\mathbf{a}_{\theta_i}^H\mathbf{X}^H\right)$ calculations is $|\hat{\Theta}|^2N$. Moreover, the total number of complex multiplication required to compute all $\mathbf{S}_{\tau}\mathbf{X}\mathbf{a}_{\theta_i}\mathbf{a}_{\theta_i}^H$, $\mathbf{S}_{\tau}^T\mathbf{X}\mathbf{a}_{\theta_i}\mathbf{a}_{\theta_i}^H $, $\mathbf{S}_{\tau}\mathbf{X}\mathbf{a}_{\theta_j}\mathbf{a}_{\theta_i}^H $ and $\mathbf{S}_{\tau}^T\mathbf{X}\mathbf{a}_{\theta_j}\mathbf{a}_{\theta_i}^H $ is $|\hat{\Theta}|^2MN$. Additionally, $\nabla_{\mathbf{X}} P(\mathbf{X})$ requires $|\mathcal{D}|$ calculations of $\nabla_{\mathbf{X}} P_{\tau}(\mathbf{X})$. Here, $|\mathcal{D}|$ is the size of $|\mathcal{D}|$ set.

As mentioned above, we have the following proposition:
\begin{proposition}
The total cost in each UM-GD iteration is roughly $\mathcal{O}\left(M^4+2M^2N+|\mathcal{D}| |\hat{\Theta}|^2MN\right)$.
\end{proposition}

\subsubsection{UM-SVRG}
In the UM-SVRG algorithm for the large-scale case, the Euclidean gradients that need to be computed in each iteration include $\nabla_{\alpha} f_{i_q^l}\left(\alpha, \mathbf{X}\right)$, $\nabla_{\mathbf{X}} e_{\theta_q^l}(\alpha, \mathbf{X})$ and $\nabla_{\mathbf{X}} P_{\tau_q^l}(\mathbf{X})$.

For $\nabla_{\alpha} f_{i_q^l}\left(\alpha, \mathbf{X}\right)$, we transform \eqref{stochastic-gradient-alpha} into
\begin{equation}\label{stochastic-gradient-alpha-analysis}
    \nabla_{\alpha} f_{i_q^l}\left(\alpha, \mathbf{X}\right)=2\left(\!\alpha \left|\bar{P}_{\theta_q^l}\right|^2-\bar{P}_{\theta_q^l}\|\mathbf{X}\mathbf{a}_{\theta_q^l}\|_2^2\right).
\end{equation}

It means that if takes $MN+N$ complex multiplications to compute \eqref{stochastic-gradient-alpha-analysis}.

For $\nabla_{\mathbf{X}} e_{\theta_q^l}(\alpha, \mathbf{X})$, we can transform \eqref{stochastic-gradient-f1-X} into
\begin{equation}  \label{f1_gradient_analysis}
\nabla_{\mathbf{X}} e_{\theta_q^l}(\alpha, \mathbf{X})=2\left(\|\mathbf{X}\mathbf{a}_{\theta_q^l}\|_2^2\mathbf{X}\mathbf{a}_{\theta_q^l}\mathbf{a}_{\theta_q^l}^H-\alpha\bar{P}_{\theta_q^l}\mathbf{X}\mathbf{a}_{\theta_q^l}\mathbf{a}_{\theta_q^l}^H\right),
\end{equation}
where $\|\mathbf{X}\mathbf{a}_{\theta_q^l}\|_2^2$ and $\mathbf{X}\mathbf{a}_{\theta_q^l}$ can be obtained when calculating $\nabla_{\alpha} f_{i_q^l}\left(\alpha, \mathbf{X}\right)$.
Therefore, it means that it takes $MN$ complex multiplications to compute $\nabla_{\mathbf{X}} e_{\theta_q^l}(\alpha, \mathbf{X})$.

For $\nabla_{\mathbf{X}} P_{\tau_q^l}(\mathbf{X})$, the number of complex multiplications required is the same as $\nabla_{\mathbf{X}} P_{\tau}(\mathbf{X})$ in the UM-GD algorithm.

Then, we have the following proposition
\begin{proposition}
The total cost in each UM-SVRG iteration is roughly $\mathcal{O}\left(|\hat{\Theta}|^2MN\right)$.
\end{proposition}

Then, we analyze the advantages of the UM-GD and UM-SVRG algorithms in the normal-scale and large-scale cases, respectively.

In the normal-scale case, $M$, $N$, and $|\mathcal{D}|$ are relatively small. Without the merging operation in \eqref{Euclidean-gradient-f1-X}, it is easy to obtain that the total  cost in each UM-GD iteration is roughly $\mathcal{O}\left(2|\Theta|MN+|\mathcal{D}| |\hat{\Theta}|^2MN\right)$. In practice, the number of elements in the set of spatial angles is large, i.e., $2|\Theta|MN$ is much greater than $M^4+2M^2N$. Therefore, the UM-GD algorithm we designed can significantly reduce the computational complexity of the algorithm.

In the large-scale case, the original UM-GD algorithm requires storing the $\sum_{\theta \in \Theta}\mathbf{A}_{\theta}\mathbf{A}_{\theta}^H$ in \eqref{Euclidean-gradient-f1-X}, which has a dimension of $M^2 \times M^2$, posing a significant challenge to the device's storage capacity. The UM-SVRG algorithm we designed not only addresses this issue but also significantly reduces the computational complexity.

In summary, the proposed UM-GD and UM-SVRG algorithms achieve low complexity in the normal-scale and large-scale cases, respectively.

\begin{figure}[htbp]
\centering
  \centerline{\includegraphics[scale=0.5]{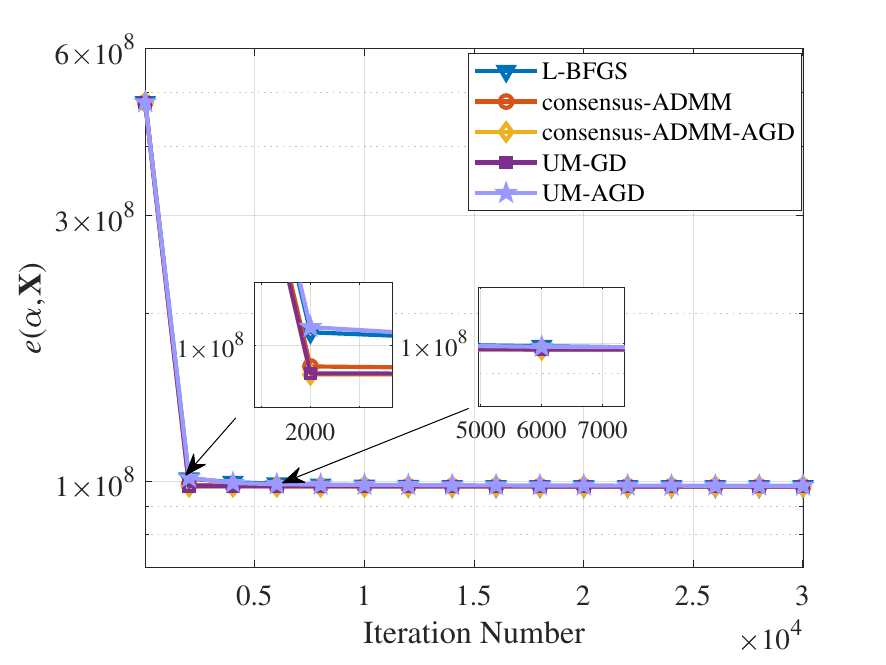}}
\centerline{(a) $e(\alpha,\mathbf{X})$ versus iteration number. }
  \centering
  \centerline{\includegraphics[scale=0.5]{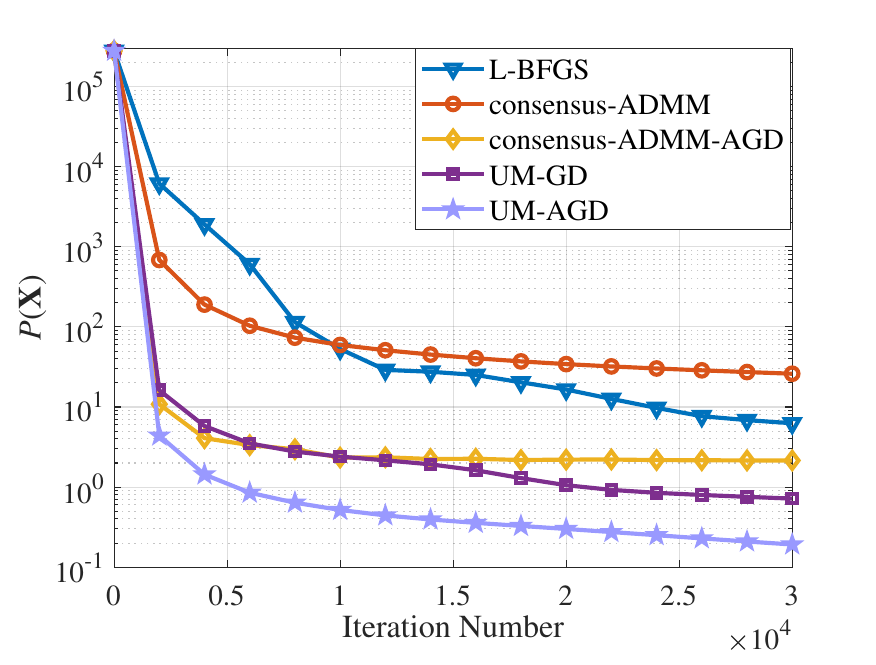}}
  \centerline{(b) $ P(\mathbf{X})$ versus iteration number.} \medskip
\caption{ Comparisons of convergence performance with $M=8,N=64,\mathcal{D}=[0,16]$.}
\label{convergence_normal_scale}
\end{figure}

\section{Simulation Results}
In this section, numerical results are presented to show the performance of the proposed MIMO radar beampattern design algorithms. Drawing from the parameter settings in existing works\cite{consensus-ADMM}, we consider the number of antennas $M = 8$ and waveform length $N = 64$ as the normal-scale case, and $M = 128$ and $N= 1024$ as the large-scale case. The set of the spacial angles covers $(-90^\circ, 90^\circ)$ with spacing $0.1^\circ$. The weight $w_c = 25$. The desired beampattern is
\begin{equation}\label{desied beampattern}
\begin{split}
\bar{P}(\theta)=\left\{
           \begin{array}{ll}
             1, & \theta\in[\theta_i-10^\circ,\theta_i+10^\circ],~i=1,2, \\
             0, & {\rm otherwise},
           \end{array}
         \right.
\end{split}
\end{equation}
where the desired angles are $\theta_1 = -40^\circ$ and $\theta_2 = 30^\circ$. In what follows, we verify the convergence of the proposed algorithms through simulations in both normal-scale and large-scale cases. Then, we demonstrate the performance advantages of the waveform designed in this work in terms of beampattern matching and correlation characteristics.

\begin{figure}[htbp]
\centering
  \centerline{\includegraphics[scale=0.48]{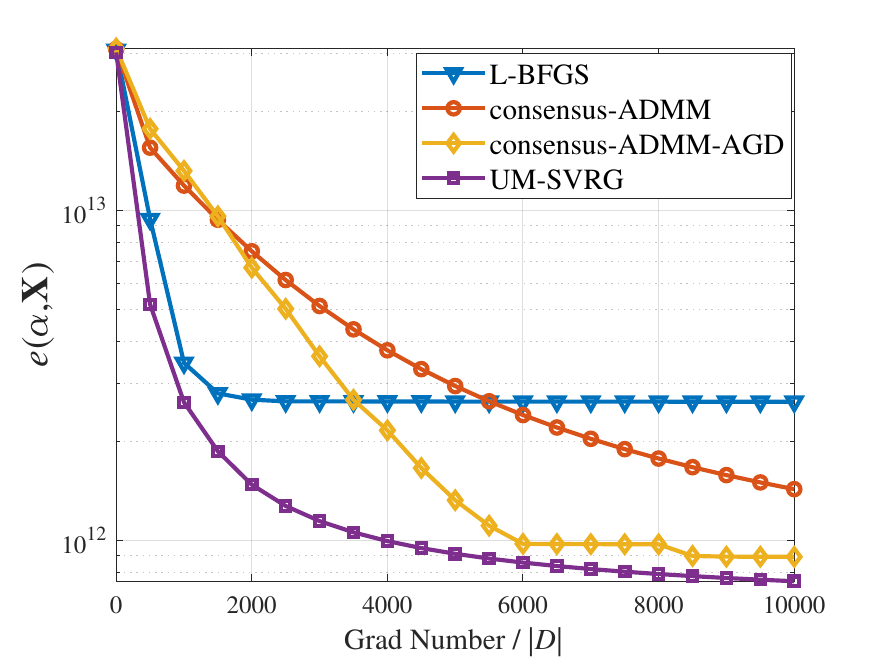}}
\centerline{(a) $e(\alpha,\mathbf{X})$ versus $\text{Grad Number}/|\mathcal{D}|$. }
  \centering
  \centerline{\includegraphics[scale=0.48]{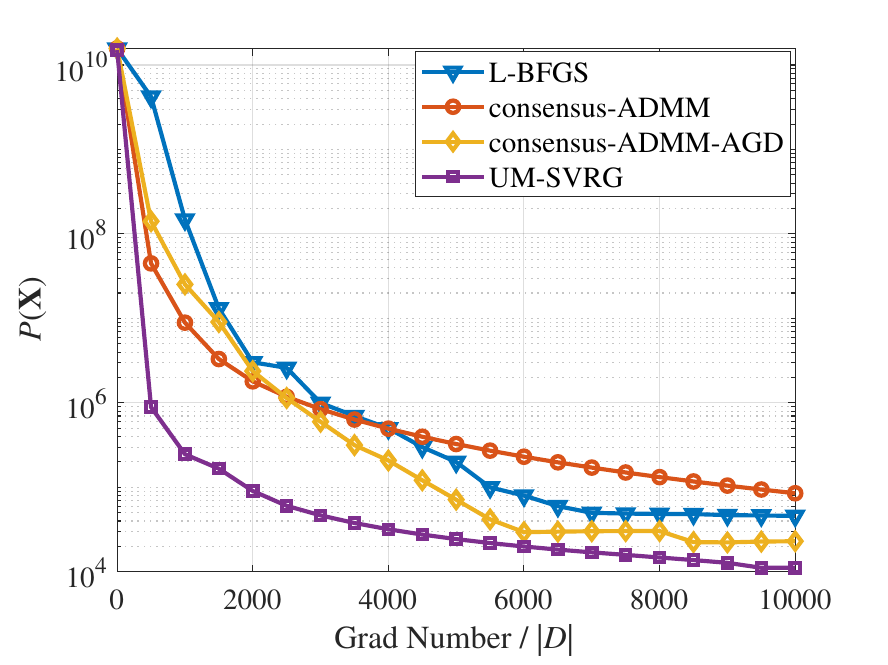}}
  \centerline{(b) $ P(\mathbf{X})$ versus $\text{Grad Number}/|\mathcal{D}|$.} \medskip
\caption{ Comparisons of convergence performance with $M=128,N=1024,\mathcal{D}=[0,256]$.}
\label{convergence_large_scale}
\end{figure}
Fig.~\ref{convergence_normal_scale} plots the performance curves of objective functions $e\left(\alpha, \mathbf{X}\right)$ and $P\left(\mathbf{X}\right)$ versus the iteration number for our proposed UM-GD and UM-AGD algorithms, as well as three other advanced algorithms, L-BFGS\cite{L-BFGS}, consensus-ADMM, and consensus-ADMM-AGD\cite{consensus-ADMM}, in the normal-scale case. It can be observed that the magnitudes of the objective function values for $e\left(\alpha, \mathbf{X}\right)$ and $P\left(\mathbf{X}\right)$ differ significantly. As a result, the performance curve of $e\left(\alpha, \mathbf{X}\right)+P\left(\mathbf{X}\right)$ fails to reflect the performance differences effectively. In the figure, all algorithms exhibit the same convergence performance on $e\left(\alpha, \mathbf{X}\right)$, while there are significant differences in performance on $P\left(\mathbf{X}\right)$. It can be observed that the proposed UM-AGD algorithm outperforms the other algorithms, while the UM-GD algorithm performs comparably to, or slightly better than, the best existing algorithm, consensus-ADMM-AGD. Moreover, as consensus-ADMM-AGD is a heuristic algorithm, the proposed algorithms have theoretical advantages over it.
On the other hand, the figure also shows that when the objective function $P(\mathbf{X})$ decreases to $10^0$, the UM-AGD algorithm reduces the required number of iterations by one order of magnitude compared to other existing algorithms, further demonstrating the superiority of the proposed algorithms.

Fig.~\ref{convergence_large_scale} plots the performance curves of objective functions $e\left(\alpha, \mathbf{X}\right)$ and $P\left(\mathbf{X}\right)$ versus the normalized gradient computation count $\text{Grad Number}/|\mathcal{D}|$ for our proposed UM-SVRG algorithm, as well as three other advanced algorithms, in the large-scale case. Since UM-SVRG requires only one sample gradient computation per iteration, while other algorithms need to compute all sample gradients, we define the normalized gradient computation count as the horizontal axis. For other algorithms, $\text{Grad Number}/|\mathcal{D}|$ is equvalent to the number of iterations. From the figure, it can be observed that our proposed UM-SVRG algorithm outperforms other algorithms in terms of both $e(\alpha,\mathbf{X})$ and $P(\mathbf{X})$. Furthermore, for a single normalized gradient computation, the computational complexity of UM-SVRG is equivalent to $\mathcal{O}\left(|\mathcal{D}| |\hat{\Theta}|^2MN\right)$, which is significantly lower than the computational complexity $\mathcal{O}(M^4+3M^2N+3|\mathcal{D}| |\hat{\Theta}|^2MN)$ of consensus-ADMM algorithm. This further highlights the advantages of the proposed UM-SVRG algorithm in terms of low complexity, superior performance, and theoretical convergence guarantee. It is also worth noting that although the UM-SVRG algorithm has low complexity, it does not demonstrate its advantage when the summation term is small, and therefore, it has no advantage in the normal-scale case.

\begin{figure}[t]
    \centering
     \includegraphics[scale = 0.41]{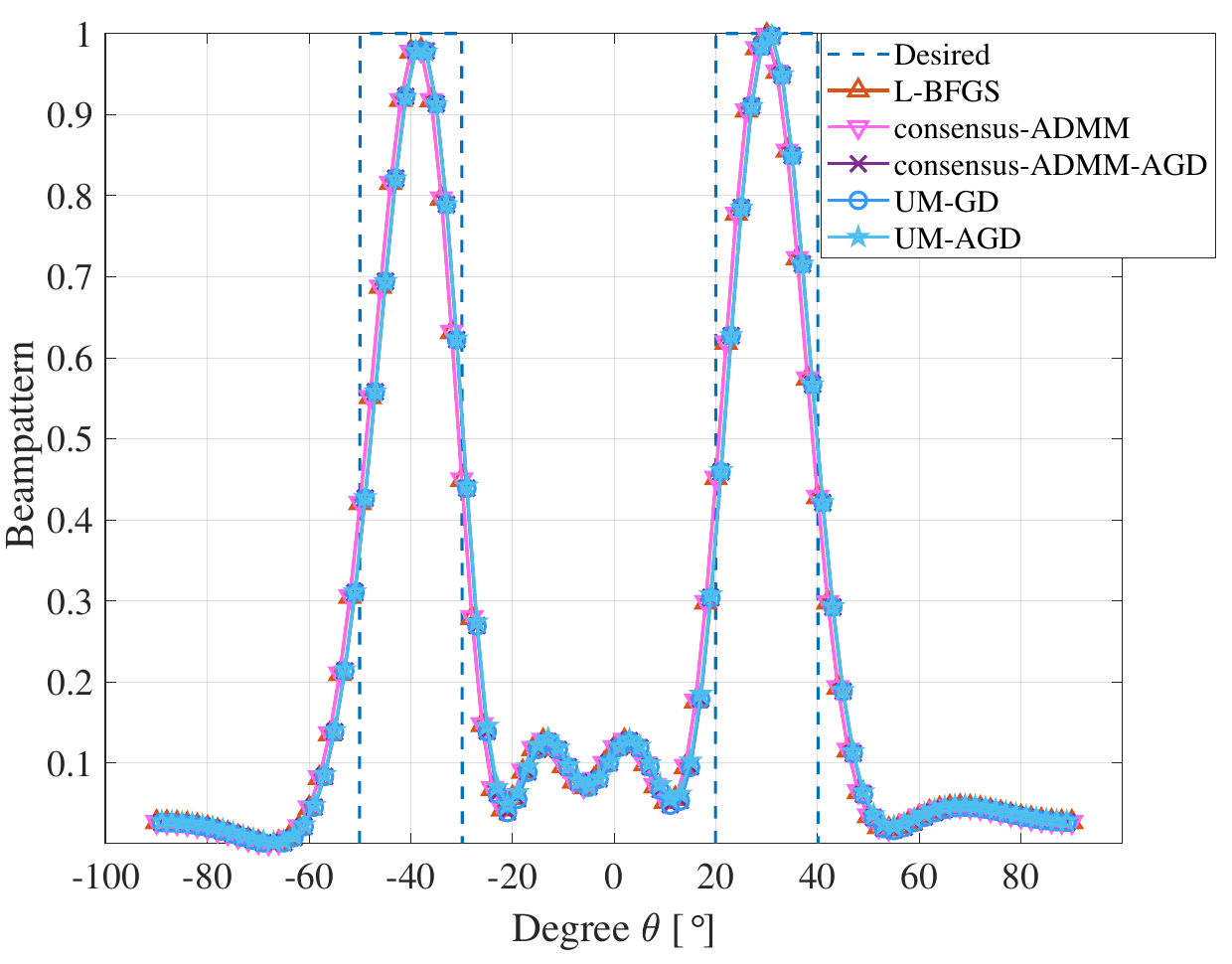}
    \caption{Comparison of synthesized beampattern with $M = 8$, $N = 64$.}
    \label{fig-beampattern-64}
\end{figure}
\begin{figure}[t]
    \centering
     \includegraphics[scale = 0.41]{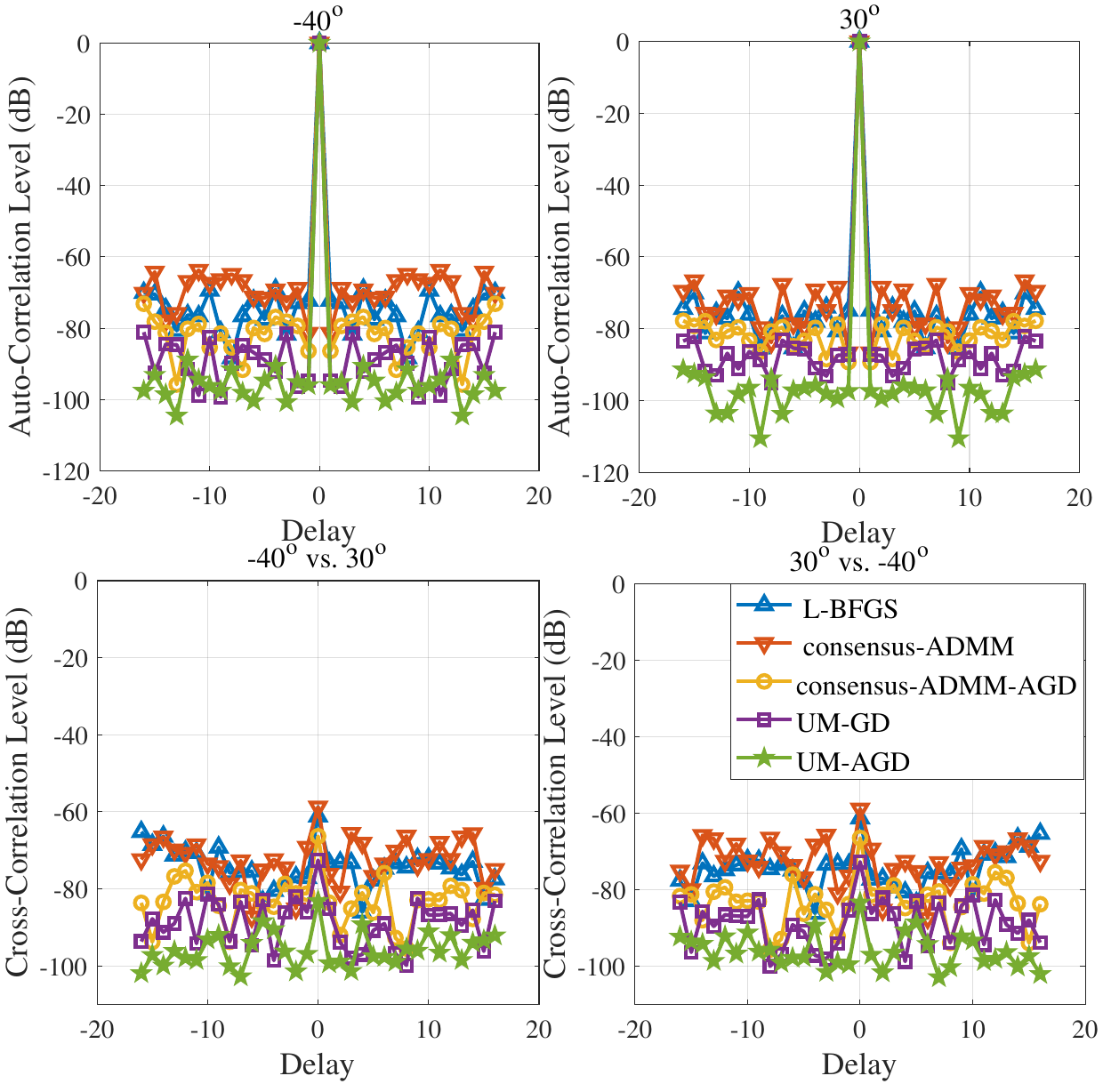}
    \caption{Comparison of correlation characteristics for interval $[0, 16]$ with $M = 8$, $N = 64$.}
    \label{fig-correlation-64}
\end{figure}
\begin{figure}[t]
    \centering
    \centerline{\includegraphics[scale=0.43]{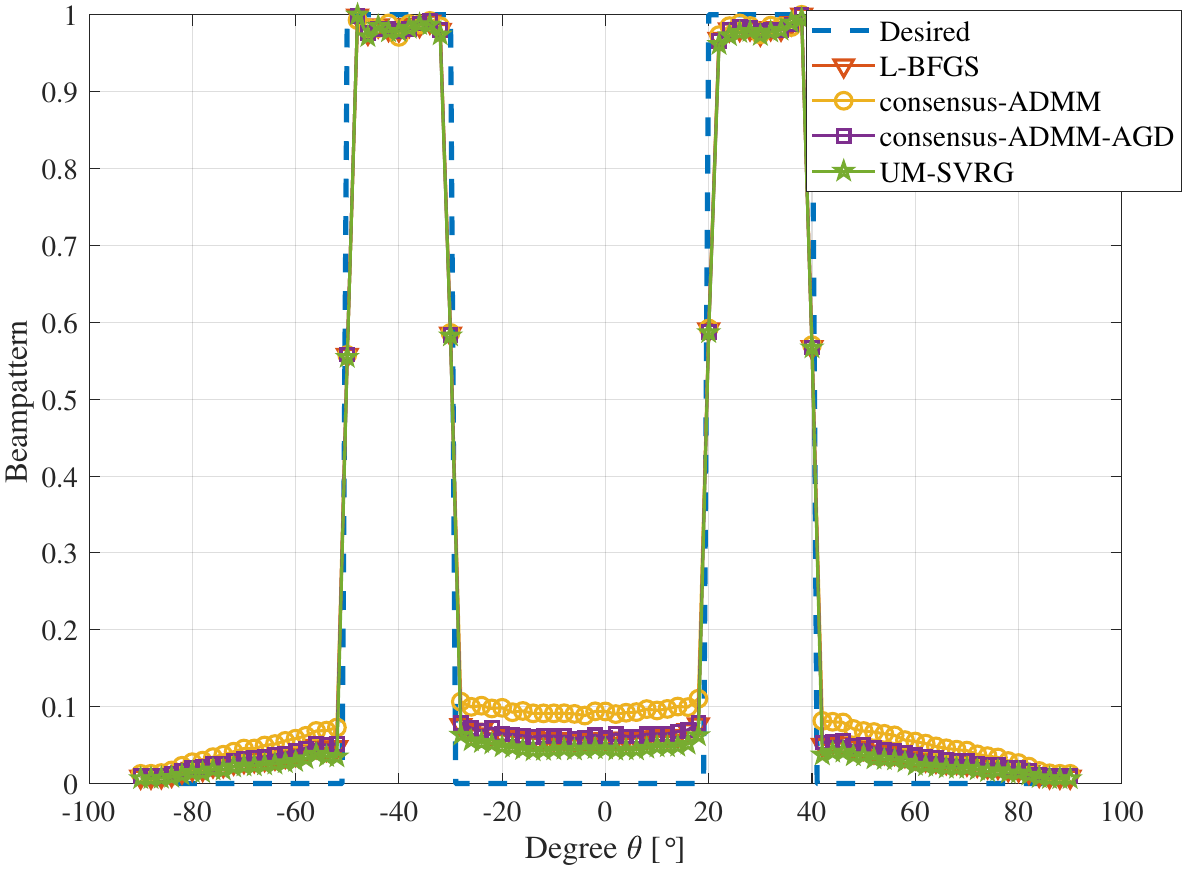}}
    \caption{Comparison of synthesized beampattern with $M = 128$, $N = 1024$.}
    \label{fig-beampattern-1024}
\end{figure}
\begin{figure}[t]
    \centering
     \includegraphics[scale = 0.43]{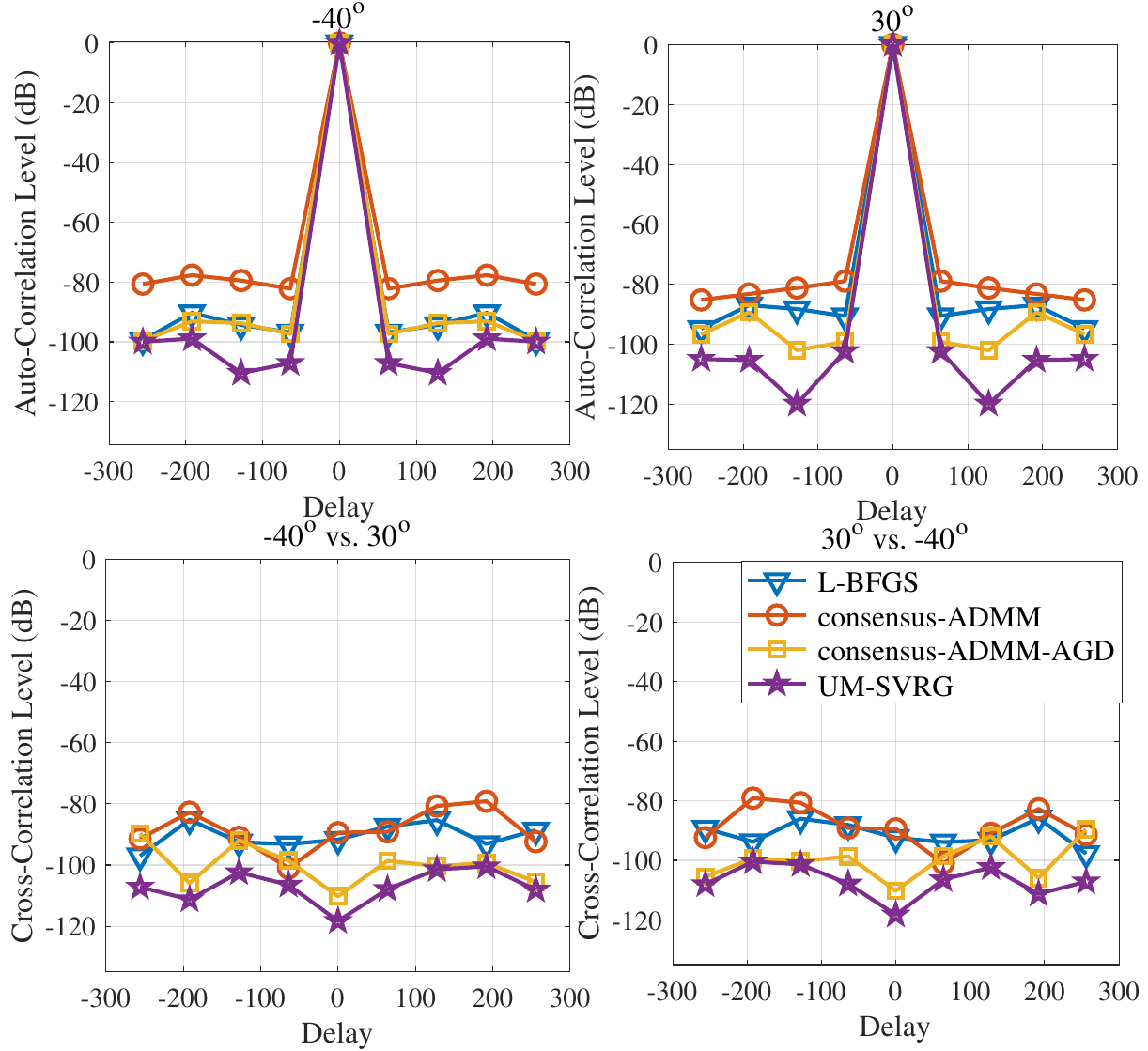}
    \caption{Comparison of correlation characteristics for interval $[0, 256]$ with $M = 128$, $N = 1024$.}
    \label{fig-correlation-1024}
\end{figure}
Fig.~\ref{fig-beampattern-64} and Fig.~\ref{fig-correlation-64} respectively show the synthesized spatial beampatterns and normalized spatial correlation levels achieved by our proposed UM-GD algorithm, UM-AGD algorithm, and three other algorithms in the normal-scale case. The normalized spatial correlation level is defined as
\begin{equation}\label{eq_autocorr}
C_{\theta_i,\theta_j,\tau}=10\log_{10}\frac{|P_{\theta_i,\theta_j,\tau} |}{\max\{|P_{\theta_i,\theta_i,0}|,|P_{\theta_j,\theta_j,0}|\}},
\end{equation}
where $\tau \in \mathcal{D}$ and $\theta_i, \theta_j \in \Theta$. From the figure, it can be observed that all the approaches can match the desired spatial beampattern very well. Moreover, our proposed UM-GD and UM-AGD algorithms outperform the other three algorithms in terms of correlation characteristics, with the UM-AGD algorithm enjoying the best auto/cross-correlation characteristics.

Fig.~\ref{fig-beampattern-1024} and Fig.~\ref{fig-correlation-1024} respectively show the synthesized spatial beampatterns and normalized spatial correlation levels in the large-scale case.  It can be observed that all the approaches can match the desired spatial beampattern very well. Additionally, compared to the other algorithms, our proposed UM-SVRG exhibits lower energy at undesired angles, indicating that it can achieve a better fitting of the beampattern.  Additionally, we can also see that the proposed UM-SVRG approach enjoys the best auto/cross-correlation characteristics. By combing Fig.~\ref{fig-correlation-64} and Fig.~\ref{fig-correlation-1024}, we can observe that the waveforms designed by our algorithm achieve correlation characteristics as low as $-100$dB in both cases, further indicating that the waveform designed by the proposed algorithm have stronger capabilities for anti-clutter interference.

\section{Conclusion}
In this paper, we focus on designing unimodular waveforms with good correlation properties for normal/large-scale MIMO radar systems. By analyzing and exploiting the inherent structure of the design problem, we embed the non-convex unimodular constraint into the search space to construct the corresponding UM manifold, transform the original problem into an unconstrained problem on the manifold. Then, for the normal-scale case, we customize a low complexity UM-GD algorithm and its accelerated version, the UM-AGD algorithm, to solve the problem.
For the large-scale case, we customize the UM-SVRG algorithm based on the structure of the objective function, significantly reducing the computational complexity of solving the problem.
We prove that the proposed three methods are theoretically convergent when appropriate parameters are chosen.
Simulation results show that our proposed algorithms achieve superior performance in both the normal-scale and large-scale cases.

\appendices
\section{Proof of Proposition \ref{UM-manifold}}\label{proof-UM-manifold}
We construct
\begin{equation}\label{operator-construct}
F: \mathbb{C}^{N \times M} \rightarrow \mathbb{R}^{N \times M}: \mathbf{X} \rightarrow \operatorname{Re}\left(\mathbf{X} \circ \mathbf{X}^*\right) - \mathbf{1}_{N \times M},
\end{equation}
where $\mathbf{1}_{N \times M}$ is an $N \times M$ matrix with all elements equal to $1$.

According to \eqref{operator-construct}, $\operatorname{dim}\left(\mathbb{C}^{N \times M}\right)=2MN$ and $\operatorname{dim}\left(\mathbb{R}^{N \times M}\right)=MN$, it follows that $\operatorname{dim}\left(\mathbb{C}^{N \times M}\right)>\operatorname{dim}\left(\mathbb{R}^{N \times M}\right)$. Moreover, it is easy to conclude that $F$ is smooth at $\mathbf{X}$.

It is clear that $\operatorname{UM}\left(N,M\right)=F^{-1}\left(\mathbf{0}_{N \times M}\right)$ and we have
\begin{equation}
\operatorname{D}F(\mathbf{X})[\mathbf{Z}]=\operatorname{Re}\left(\mathbf{Z} \circ \mathbf{X}^*\right).
\end{equation}

It is easy to see that for all $\hat{\mathbf{Z}} \in \mathbb{R}^{N \times M}$, there exists $\mathbf{Z} = \hat{\mathbf{Z}} \circ \mathbf{X} \in \mathbb{C}^{N \times M}$ such that $\operatorname{D}F(\mathbf{X})[\mathbf{Z}]=\hat{\mathbf{Z}}$. This shows that $F$ is full rank. It follows from submersion theorem \cite{manifold-optimization-absil} that the set $\operatorname{UM}\left(N,M\right)$ defined in \eqref{UM-definition} is an embedded submanifold of $\mathbb{C}^{N \times M}$. Moreover, $\operatorname{dim}\left(\operatorname{UM}\left(N,M\right)\right)=\operatorname{dim}\left(\mathbb{C}^{N \times M}\right)-\operatorname{dim}\left(\mathbb{R}^{N \times M}\right) = 2MN-MN =MN$.$\hfill\blacksquare$

\section{Proof of Lemma \ref{lemma-tangent-space} and Lemma \ref{lemma-Riemannian-metric}}\label{proof-tangent}
According to Appendix \ref{proof-UM-manifold}, it can be concluded that $\operatorname{UM}\left(N,M\right)$ is defined as a level set of a constant-rank function (specifically in the form of \eqref{operator-construct}), we have
\begin{equation}\label{tangent-kernel}
T_{\mathbf{X}}\operatorname{UM}\left(N,M\right)=\operatorname{ker}\left(\operatorname{D}F\left(\mathbf{X}\right)\right).
\end{equation}

From \eqref{operator-construct} and \eqref{tangent-kernel}, it is easy to derive \eqref{tangent-space}.

As we know, manifold $\mathbb{C}^{N \times M}$ is endowed with a Riemannian metric. Then, we would expect that manifold $\operatorname{UM}\left(N,M\right)$ generated from $\mathbb{C}^{N \times M}$ can inherit a Riemannian metric in a natural way.

Since every tangent space $T_{\mathbf{X}}\operatorname{UM}\left(N,M\right)$ can be regarded as a subspace of $T_{\mathbf{X}}\mathbb{C}^{N \times M}$, the Riemannian metric $\bar{g}$ of $T_{\mathbf{X}}\mathbb{C}^{N \times M}$ induces a Rimannian metric $g$ on $\operatorname{UM}\left(N,M\right)$ according
\begin{equation}
g_{\mathbf{X}}\left(\boldsymbol{\xi}_{\mathbf{X}},\boldsymbol{\eta}_{\mathbf{X}}\right)=\bar{g}_{\mathbf{X}}\left(\boldsymbol{\xi}_{\mathbf{X}},\boldsymbol{\eta}_{\mathbf{X}}\right),
\end{equation}
where $\boldsymbol{\xi}_{\mathbf{X}}, \boldsymbol{\eta}_{\mathbf{X}} \in T_{\mathbf{X}}\operatorname{UM}\left(N,M\right)$, and $\boldsymbol{\xi}_{\mathbf{X}}, \boldsymbol{\eta}_{\mathbf{X}}$ can also be viewed as elements of  $T_{\mathbf{X}}\mathbb{C}^{N \times M}$.

Since the tangent space $T_{\mathbf{X}}\mathbb{C}^{N \times M}$ is the same as $\mathbb{C}^{N \times M}$, the inner product on $\mathbb{C}^{N \times M}$ is \cite{complex-analysis}
\begin{equation}
\left\langle \boldsymbol{\xi}_{\mathbf{X}}, \boldsymbol{\eta}_\mathbf{X}\right\rangle_{\mathbf{X}}=\operatorname{Re}\left(\operatorname{Tr}\left(\boldsymbol{\xi}_{\mathbf{X}}^H\boldsymbol{\eta}_{\mathbf{X}}\right)\right),
\end{equation}
where $\boldsymbol{\xi}_{\mathbf{X}}, \boldsymbol{\eta}_{\mathbf{X}} \in \mathbb{C}^{N \times M}$.

Then, we can get \eqref{Riemannian-metric}.$\hfill\blacksquare$

\section{Proof of Lemma \ref{Riemannian-gradient}}\label{proof-Riemannian-gradient}
Any element $\nabla f\left(\mathbf{X}\right) \in \mathbb{C}^{N \times M}$ can be uniquely decomposed into the sum of an element of $T_{\mathbf{X}}\operatorname{UM}\left(N,M\right)$ and an element of $\left(T_{\mathbf{X}}\operatorname{UM}\left(N,M\right)\right)^{\perp}$. That is
\begin{equation}
\nabla f\left(\mathbf{X}\right)=\operatorname{P}_{\mathbf{X}}\nabla f\left(\mathbf{X}\right)+\operatorname{P}_{\mathbf{X}}^{\perp}\nabla f\left(\mathbf{X}\right),
\end{equation}
where $\operatorname{P}_{\mathbf{X}}$ denotes the orthogonal projection onto $T_{\mathbf{X}}\operatorname{UM}\left(N,M\right)$ and $\operatorname{P}_{\mathbf{X}}^{\perp}$ denotes the orthogonal projection onto $\left(T_{\mathbf{X}}\operatorname{UM}\left(N,M\right)\right)^{\perp}$.

Since $\operatorname{P}_{\mathbf{X}}\nabla f\left(\mathbf{X}\right)$ belongs to  $T_{\mathbf{X}}\operatorname{UM}\left(N,M\right)$ and satisfies that $\left\langle\operatorname{P}_{\mathbf{X}}\nabla f\left(\mathbf{X}\right),\boldsymbol{\eta}_{\mathbf{X}}\right\rangle \!=\! \operatorname{D}f\left(\mathbf{X}\right)\left[\boldsymbol{\eta}_{\mathbf{X}}\right]$ for all $\boldsymbol{\eta}_{\mathbf{X}} \in T_{\mathbf{X}}\operatorname{UM}\left(N,M\right)$, we can get the Riemannian gradient $\operatorname{grad}f\left(\mathbf{X}\right)$ of $f$ is
\begin{equation}\label{Riemannian-gradient-projection}
\operatorname{grad}f\left(\mathbf{X}\right)=\operatorname{P}_{\mathbf{X}}\nabla f\left(\mathbf{X}\right).
\end{equation}

Next, we first derive the explicit expression for $\left(T_{\mathbf{X}}\operatorname{UM}\left(N,M\right)\right)^{\perp}$, and then provide the specific computational expression for \eqref{Riemannian-gradient-projection}.

The orthogonal complement $\left(T_{\mathbf{X}}\operatorname{UM}\left(N,M\right)\right)^{\perp}$ of $T_{\mathbf{X}}\operatorname{UM}\left(N,M\right)$ is denoted by
\begin{equation}
\begin{aligned}
&\left(T_{\mathbf{X}}\operatorname{UM}\left(N,M\right)\right)^{\perp} =\\
&\left\{\boldsymbol{\zeta} \in \mathbb{C}^{N \times M}: \bar{g}_{\mathbf{X}}\left(\boldsymbol{\zeta},\boldsymbol{\xi}_{\mathbf{X}}\right)=0 \ \text{for all} \  \boldsymbol{\xi}_{\mathbf{X}} \in T_{\mathbf{X}}\operatorname{UM}\left(N,M\right)\right\}.
\end{aligned}
\end{equation}

It is easy to get
\begin{equation}
\left(T_{\mathbf{X}}\operatorname{UM}\left(N,M\right)\right)^{\perp} =\left\{\mathbf{X} \circ \mathbf{A}: \mathbf{A} \in \mathbb{R}^{N \times M} \right\}.
\end{equation}

Then, for any $\nabla f \left(\mathbf{X}\right) \in \mathbb{C}^{N \times M}$
\begin{equation}
\nabla f \left(\mathbf{X}\right)-\operatorname{P}_{\mathbf{X}}\left(\nabla f \left(\mathbf{X}\right)\right) = \operatorname{P}_{\mathbf{X}}^{\perp}\left(\nabla f \left(\mathbf{X}\right)\right) = \mathbf{X} \circ \mathbf{A}.
\end{equation}

According to \eqref{tangent-space}
\begin{equation}
\operatorname{Re}\left(\left(\nabla f \left(\mathbf{X}\right)-\mathbf{X} \circ \mathbf{A}\right)\circ \mathbf{X}^*\right)=0.
\end{equation}

Then, $\mathbf{A}$ is obtained as
\begin{equation}
\mathbf{A}=\operatorname{Re}\left(\nabla f \left(\mathbf{X}\right)\circ \mathbf{X}^*\right).
\end{equation}

Thus, we obtain \eqref{Riemannian-gradient-operator}.$\hfill\blacksquare$

\section{Proof of Lemma \ref{Retraction}}\label{proof-Retraction}
According to \cite{manifold-optimization-absil}, we have the following proposition.
\begin{proposition}\label{retraction-proposition}
Let $\mathcal{N}$ be an abstract manifold such that $\operatorname{dim}\left(\operatorname{UM}\left(N,M\right)\right)+\operatorname{dim}\left(\mathcal{N}\right)
=\operatorname{dim}\left({\mathbb{C}^{N \times M}}\right)$. Assume that there is a diffeomorphism
\begin{equation}
\phi: \operatorname{UM}\left(N,M\right) \times \mathcal{N} \rightarrow \mathcal{E}_*: \left(\mathbf{X},\mathbf{B}\right) \rightarrow \phi \left(\mathbf{X},\mathbf{B}\right),
\end{equation}
where $\mathcal{E}_*$ is an open subset of $\mathbb{C}^{N \times M}$. The mapping
\begin{equation}
R_{\mathbf{X}}\left(\boldsymbol{\xi}_{\mathbf{X}}\right):=\pi_1\left(\phi^{-1}\left(\mathbf{X}+\boldsymbol{\xi}_{\mathbf{X}}\right)\right),
\end{equation}
where $\pi_1: \operatorname{UM}\left(N,M\right) \times \mathcal{N} \rightarrow \operatorname{UM}\left(N,M\right): \left(\mathbf{X},\mathbf{B}\right) \rightarrow \mathbf{X}$, defines a retraction on $\operatorname{UM}\left(N,M\right)$.
\end{proposition}

We construct an abstract manifold $\mathcal{N}$ as $\mathbb{R}^{N \times M}_+$, which is a $N \times M$ real matrix where each element is greater than $0$. It is easy to obtain $\operatorname{dim}\left(\mathbb{R}^{N \times M}_+\right)=MN$, satisfying $\operatorname{dim}\left(\operatorname{UM}\left(N,M\right)\right)+\operatorname{dim}\left(\mathbb{R}^{N \times M}_+\right)
=\operatorname{dim}\left({\mathbb{C}^{N \times M}}\right)$.

Then, we construct the mapping
\begin{equation}
\phi: \operatorname{UM}\left(N,M\right) \times \mathbb{R}^{N \times M}_+ \rightarrow \mathbb{C}^{N \times M}_*: \left(\mathbf{X},\mathbf{B}\right) \rightarrow \mathbf{X} \circ \mathbf{B},
\end{equation}
where $\mathbb{C}^{N \times M}_*$ is an $N \times M$ complex matrix with no zero elements.

It is easy to see that the mapping $\phi$ is a bijection such that $\phi$ and its inverse $\phi^{-1}$ are both smooth, hence $\phi$ is a diffeomorphism.

According to  Proposition \ref{retraction-proposition}, we obtain \eqref{retraction-express}.$\hfill\blacksquare$

\section{Proof of Euclidean Gradient}\label{proof-euclidean-gradient}
First, we analysis $\nabla_\alpha e(\alpha, \mathbf{X})$. Since $e(\alpha, \mathbf{X})$ is a real-valued function of the real number $\alpha$, we have
\begin{equation}
\begin{aligned}
\nabla_\alpha e(\alpha, \mathbf{X})& = 2\sum_{\theta \in \Theta}\left(\alpha \bar{P}_{\theta}-\mathbf{a}_{\theta}^H\mathbf{X}_H\mathbf{X}\mathbf{a}_{\theta}\right)\bar{P}_{\theta}\\
& = 2\left(\alpha \sum_{\theta \in \Theta}\left|\bar{P}_\theta\right|^2-\operatorname{Tr}\left(\mathbf{X}^H \mathbf{X}\right) \sum_{\theta \in \Theta} \bar{P}_\theta \mathbf{a}_\theta \mathbf{a}_\theta^H\right).
\end{aligned}
\end{equation}

Additionally, $e(\alpha, \mathbf{X})$ and $\alpha \bar{P}_\theta-\mathbf{a}_\theta^H \mathbf{X}^H \mathbf{X} \mathbf{a}_\theta$ are real-valued functions of the complex variable $\mathbf{X}$. From \cite{matrix-cookbook}, we have
\begin{equation} \label{Euclidean-gradient-second-part}
\begin{aligned}
&\nabla_\mathbf{X} e(\alpha, \mathbf{X})\\
& = 2\left(-\alpha \mathbf{X}\sum_{\theta \in \Theta}\left(\bar{P}_{\theta}\mathbf{a}_{\theta}\mathbf{a}_{\theta}^H\right)+\sum_{\theta \in \Theta}\left(\mathbf{a}_{\theta}^H\mathbf{X}^H\mathbf{X}\mathbf{a}_{\theta}\mathbf{X}\mathbf{a}_{\theta}\mathbf{a}_{\theta}^H\right)\right).
\end{aligned}
\end{equation}

The second term $\sum_{\theta \in \Theta}\left(\mathbf{a}_{\theta}^H\mathbf{X}^H\mathbf{X}\mathbf{a}_{\theta}\mathbf{X}\mathbf{a}_{\theta}\mathbf{a}_{\theta}^H\right)$ in \eqref{Euclidean-gradient-second-part} remains in the form of a polynomial summation regarding $\mathbf{X}$, which requires further simplification. Then, we have
\begin{equation}
\begin{aligned}
\operatorname{vec}\left(\mathbf{a}_{\theta}^H\mathbf{X}^H\mathbf{X}\mathbf{a}_{\theta}\mathbf{X}\mathbf{a}_{\theta}\mathbf{a}_{\theta}^H\right)
&=\mathbf{a}_{\theta}^H\mathbf{X}^H\mathbf{X}\mathbf{a}_{\theta}\operatorname{vec}\left(\mathbf{X}\mathbf{a}_{\theta}\mathbf{a}_{\theta}^H\right)\\
& =  \left(\mathbf{I}_m \otimes \mathbf{X} \right)\mathbf{A}_{\theta}\mathbf{A}_{\theta}^H\operatorname{vec}\left(\mathbf{X}^H\mathbf{X}\right).
\end{aligned}
\end{equation}
\begin{equation}
\begin{aligned}
&\sum_{\theta \in \Theta}\left(\mathbf{a}_{\theta}^H\mathbf{X}^H\mathbf{X}\mathbf{a}_{\theta}\mathbf{X}\mathbf{a}_{\theta}\mathbf{a}_{\theta}^H\right) \\
&= \sum_{\theta \in \Theta}\operatorname{unvec}\left(\left(\mathbf{I}_m \otimes \mathbf{X} \right)\mathbf{A}_{\theta}\mathbf{A}_{\theta}^H\operatorname{vec}\left(\mathbf{X}^H\mathbf{X}\right)\right)\\
&= \operatorname{unvec}\left(\left(\mathbf{I}_m \otimes \mathbf{X} \right)\sum_{\theta \in \Theta}\left(\mathbf{A}_{\theta}\mathbf{A}_{\theta}^H\right)\operatorname{vec}\left(\mathbf{X}^H\mathbf{X}\right)\right).
\end{aligned}
\end{equation}

Thus, we can get \eqref{Euclidean-gradient-f1-X}.

$\left|P_{\theta_i, \theta_i, \tau}\right|^2$ and $\left|P_{\theta_i, \theta_j, \tau}\right|^2$ are real-valued functions of the complex variable $\mathbf{X}$, while $P_{\theta_i, \theta_i, \tau}$ and $P_{\theta_i, \theta_j, \tau}$ are complex-valued functions of the complex variable $\mathbf{X}$. By the chain rule, we have
\begin{equation}\label{P-norm2-gradient}
\frac{\partial\left|P_{\theta_i, \theta_i, \tau}\right|^2}{\partial \mathbf{X}^*} =  P_{\theta_i, \theta_i, \tau}^*\frac{\partial P_{\theta_i, \theta_i, \tau}}{\partial \mathbf{X}^*}+P_{\theta_i, \theta_i, \tau}\frac{\partial P_{\theta_i, \theta_i, \tau}^*}{\partial \mathbf{X}^*}.
 \end{equation}

Then, we have
\begin{equation}\label{P-gradient-part1}
\frac{\partial P_{\theta_i, \theta_i, \tau}}{\partial \mathbf{X}^*} = \mathbf{S}_{\tau}\mathbf{X}\mathbf{a}_{\theta_i}\mathbf{a}_{\theta_i}^H,
\end{equation}
\begin{equation}\label{P-gradient-part2}
\frac{\partial P_{\theta_i, \theta_i, \tau}^*}{\partial \mathbf{X}^*} = \mathbf{S}_{\tau}^T\mathbf{X}\mathbf{a}_{\theta_i}\mathbf{a}_{\theta_i}^H.
\end{equation}

Substituting \eqref{P-gradient-part1} and \eqref{P-gradient-part2} into \eqref{P-norm2-gradient}, we have \eqref{p-norm2-gradient-specific1}. Similarly, we can obtain \eqref{p-norm2-gradient-specific2}.$\hfill\blacksquare$

\section{Proof of Theorem \ref{R-SD-convergence}}\label{proof-R-SD-convergence}
We have the following lemma to prove that $f\left(R_{\left(\alpha, \mathbf{X}\right)}\left(\mathbf{s}\right)\right)$ satisfies \eqref{lipschitz-continuous}, and then we will prove Theorem~\ref{R-SD-convergence}. In the subsequent proof, $\|\mathbf{S}\|_{\left(\alpha, \mathbf{X}\right)}^2 = g_{\left(\alpha, \mathbf{X}\right)}\left(\mathbf{S}, \mathbf{S}\right)$ denotes the norm based on the Riemannian metric on $T_{\left(\alpha, \mathbf{X}\right)}\mathcal{M}$.
\begin{lemma}
Consider the retraction $R_{\left(\alpha, \mathbf{X}\right)}$ on $\mathcal{M}$, a compact subset $\mathcal{K} \subseteq \mathcal{M}$ and a continuous, nonnegative function $r: \mathcal{K} \rightarrow \mathbb{R}$. The set
\begin{equation}
\mathcal{T}=\{(\left(\alpha, \mathbf{X}\right), \mathbf{S}) \in T\mathcal{M}: \mathbf{X} \in \mathcal{K} \text { and }\|\mathbf{S}\|_{\left(\alpha, \mathbf{X}\right)}^2 \leq r\left(\alpha, \mathbf{X}\right)\}
\end{equation}
is compact in the tangent bundle $T\mathcal{M}$. $f: \mathcal{M} \rightarrow \mathbb{R}$ is twice continuously differentiable. There exists a constant $L$ such that, for all $(\left(\alpha, \mathbf{X}\right), \mathbf{S}) \in \mathcal{T}$, with $\hat{f}=f \circ \mathrm{R}_x$, we have
\begin{equation}
\left|f\left(R_{\left(\alpha, \mathbf{X}\right)}\left(\mathbf{S}\right)\right)\!-\!f\left(\alpha, \mathbf{X}\right)\!-\!\langle \operatorname{grad}\! f\left(\alpha,\! \mathbf{X}\right), \mathbf{X}\rangle\right|\! \leq \! \frac{L}{2}\|\mathbf{S}\|_{\left(\alpha, \mathbf{X}\right)}^2.
\end{equation}

Then we have
\begin{equation}
f\left(R_{\left(\alpha, \mathbf{X}\right)}\left(\mathbf{S}\right)\right)\leq f\left(\alpha, \mathbf{X}\right)+\langle\operatorname{grad} f(\alpha, \mathbf{X}), \mathbf{S}\rangle+\frac{L}{2}\|\mathbf{S}\|_{\left(\alpha, \mathbf{X}\right)}^2 .
\end{equation}
\end{lemma}
{\it Proof:} The proof can be found in \cite{manifold-nicolas}.

For all $k$
\begin{equation}
f\!\left(\!\alpha_{k+1}, \!\mathbf{X}_{k+1}\!\right)\! \leq\! f\!\left(\!\alpha_k,\! \mathbf{X}_k\!\right)+\left\langle\operatorname{grad} \!\!f\!\left(\!\alpha_k, \!\mathbf{X}_k\!\right),\! \mathbf{s}_k\right\rangle\!+\frac{L}{2}\!\!\left\|\mathbf{S}_k\right\|_{\left(\alpha_k, \!\mathbf{X}_k\right)}^2.
\end{equation}
Reorganizing and using $\mathbf{s}_k=-t_k \operatorname{grad} f\left(\alpha_k, \mathbf{X}_k\right)$ reveals
\begin{equation}\label{tk-lipschitz}
f\!\left(\!\alpha_k,\! \mathbf{X}_k\!\right)\!-\!f\!\left(\!\alpha_{k+1},\! \mathbf{X}_{k+1}\!\right) \!\geq\!\left(\!\!t_k\!-\!\frac{L}{2} t_k^2\!\!\right)\!\left\|\operatorname{grad} f\!\left(\alpha_k,\! \mathbf{X}_k\right)\right\|_{\left(\alpha_k,\! \mathbf{X}_k\right)}^2,
\end{equation}
$t_k\!-\!\frac{L}{2} t_k^2$ is quadratic in $t_k$, positive between 0 and $2 / L$.

Since the value of $f$ is non-negative, it follows that $f_{low} = 0$, which makes $f\left(\alpha, \mathbf{X}\right) \geq f_{low}$ for all $\left(\alpha, \mathbf{X}\right) \in \mathcal{M}$.

For all $K>1$, we can get the inequality as follows
\begin{equation}
\begin{aligned}
f\left(\alpha_0, \mathbf{X}_0\right)\!-\!f_{low} &\geq f\left(\alpha_0, \mathbf{X}_0\right)- f\left(\alpha_K, \mathbf{X}_K\right)\\
&= \sum_{k=0}^{K-1}f\left(\alpha_k, \mathbf{X}_k\right)- f\left(\alpha_{k+1}, \mathbf{X}_{k+1}\right)\\
&\geq K \epsilon \min_{k= 0,...,K-1}\|\operatorname{grad}f\left(\alpha_k, \mathbf{X}_k\right)\|_{\left(\alpha_k, \mathbf{X}_k\right)}^2,
\end{aligned}
\end{equation}
where $\epsilon = \min\left(t_k-\frac{L}{2}t_k^2\right) $.

Taking $K \rightarrow \infty$, we can get
\begin{equation}\label{Infinite-sum-fk}
    f\left(\alpha_0, \mathbf{X}_0\right)-f_{low} \geq  \sum_{k=0}^{\infty}f\left(\alpha_k, \mathbf{X}_k\right)- f\left(\alpha_{k+1}, \mathbf{X}_{k+1}\right).
\end{equation}

According to \eqref{tk-lipschitz}, the right-hand side is a series of positive terms. Inequality \eqref{Infinite-sum-fk} implies that positive series converges, thus
\begin{equation}
\begin{aligned}
0 &= \lim_{k\rightarrow\infty}f\left(\alpha_k, \mathbf{X}_k\right)- f\left(\alpha_{k+1}, \mathbf{X}_{k+1}\right) \\
& \geq \epsilon\lim_{k\rightarrow\infty}\|\operatorname{grad}\left(\alpha_k, \mathbf{X}_k\right)\|_{\left(\alpha_k, \mathbf{X}_k\right)}^2,
\end{aligned}
\end{equation}
which confirms that $\|\operatorname{grad}\left(\alpha_k, \mathbf{X}_k\right)\|_{\left(\alpha_k, \mathbf{X}_k\right)}^2 \rightarrow 0$ and shows all accumulation points are stationary points.

\begin{remark}
For the accelerated UM-GD algorithm, since the chosen step size satisfies $\eqref{Accelerate-R-SD-stepsize-condition}$, which implies it satisfies \eqref{tk-lipschitz}, it can shown through the subsequent proof steps that the proposed UM-AGD algorithm converges to a stationary point.
\end{remark}
$\hfill\blacksquare$

\section{Proof of Theorem \ref{R-SVRG-convergence}}\label{proof-R-SVRG-convergence}
Based on \cite{topology}, it is easy to conclude that there exists a connected and compact set $\mathcal{U} \subset \mathcal{M}$ such that for any $i, j >0$,  $\left(\alpha_q^l, \mathbf{X}_q^l\right) \in \mathcal{N}$ for the element $\left(\alpha_q^l, \mathbf{X}_q^l\right)$ in the sequence $\{\left(\alpha_q^l, \mathbf{X}_q^l\right)\}$. Then, any continuous function on $\mathcal{U}$ is bounded. According to the triangle inequality, we have
\begin{equation}\label{triangle inequality R-SVRG gradient bound}
\left\|\boldsymbol{\varsigma}_{q+1}^l\right\|_{\left(\alpha_q^l, \mathbf{X}_q^l\right)} \leq C.
\end{equation}

Define the function
\begin{equation}
g\left(t ; \alpha_q^l, \mathbf{X}_q^l, \boldsymbol{\varsigma}_{q+1}^l\right)=f \circ R_{\left(\alpha_q^l, \mathbf{X}_q^l\right)}\left(-t \boldsymbol{\varsigma}_{q+1}^l\right).
\end{equation}

Hence, there exists a constant $k_1$ such that
\begin{equation}\label{smooth}
\frac{d^2}{d t^2} g\left(t ; \alpha_q^l, \mathbf{X}_q^l, \boldsymbol{\varsigma}_{q+1}^l\right) \leq 2 k_1,
\end{equation}
where $t \in\left[0, \alpha_0^1\right], \left(\alpha_q^l, \mathbf{X}_q^l\right) \in \mathcal{M}$.

According to the Taylor expansion of integral remainder, we have
\begin{equation}\label{fX decrease}
\begin{aligned}
& f\left(\alpha_{q+1}^l, \mathbf{X}_{q+1}^l\right)-f\left(\alpha_q^l, \mathbf{X}_q^l\right) \\
& =g\left(t_q^l; \alpha_q^l, \mathbf{X}_q^l, \varsigma_{q+1}^l\right)-g\left(0; \alpha_q^l, \mathbf{X}_q^l, \boldsymbol{\varsigma}_{q+1}^l\right) \\
& =-t_q^l\!\left\langle \boldsymbol{\varsigma}_{q+1}^l, \operatorname{grad}f\left(\alpha_q^l, \mathbf{X}_q^l\right)\right\rangle_{\left(\alpha_q^l, \mathbf{X}_q^l\right)}+\\
& \quad \left(t_q^l\right)^2 \int_0^1\left(1-t_1\right) g^{\prime \prime}\left(t_q^l t_1; \alpha_q^l, \mathbf{X}_q^l, \boldsymbol{\varsigma}_{q+1}^l\right) d t_1 \\
& \leq-t_q^l\left\langle\boldsymbol{\varsigma}_{q+1}^l, \operatorname{grad} f\left(\alpha_q^l, \mathbf{X}_q^l\right)\right\rangle_{\left(\alpha_q^l, \mathbf{X}_q^l\right)}+\left(t_q^l\right)^2 k_1.
\end{aligned}
\end{equation}

Since each iteration involves randomly selecting sample gradients, $f\left(\alpha_q^l, \mathbf{X}_q^l\right)$ is a stochastic process.

We construct $\mathcal{F}_q^l$ as an increasing sequence of $\sigma$-algebras \cite{measure-theory}
\begin{equation}\label{sigma algebra}
\mathcal{F}_q^l=\left\{i_1^1, i_2^1, \ldots, i_{m_1}^1, i_1^2, i_2^2, \ldots, i_{m_2}^2, i_1^l, \ldots, i_{q-1}^l\right\}.
\end{equation}

$\left(\alpha_q^l, \mathbf{X}_q^l\right)$ is calculated based on $i_1^1, i_2^1, \ldots, i_q^l$, and is measurable on $\mathcal{F}_{q+1}^l$, and $i_{q+1}^l$ is independent of $\mathcal{F}_{q+1}^l$, we have
\begin{equation}\label{expectation inner product}
\mathbb{E}\!\left[\!\left\langle\!\boldsymbol{\varsigma}_{q+1}^l,\! \operatorname{grad}\!\! f\!\left(\alpha_q^l,\! \mathbf{X}_q^l\right)\!\right\rangle_{\left(\!\alpha_q^l, \!\mathbf{x}_q^l\!\right)}\! \mid\! \mathcal{F}_{q+1}^l\!\right]\!\!=\!\!\left\|\operatorname{grad}\!\! f\!\left(\alpha_q^l, \!\mathbf{X}_q^l\right)\!\right\|_{\left(\!\alpha_q^l, \!\mathbf{X}_q^l\!\right)}^2.
\end{equation}

Then, \eqref{expectation inner product} leads to the following inequality
\begin{equation}\label{fX decrease bound}
\begin{aligned}
&\mathbb{E}\left[f\left(\alpha_{q+1}^l, \mathbf{X}_{q+1}^l\right)-f\left(\alpha_q^l, \mathbf{X}_q^l\right) \mid \mathcal{F}_{q+1}^l\right] \leq \\
&-t_q^l\left\|\operatorname{grad} f\left(\alpha_q^l, \mathbf{X}_q^l\right)\right\|_{\left(\alpha_q^l, \mathbf{X}_q^l\right)}^2+\left(t_q^l\right)^2 k_1 \leq\left(t_q^l\right)^2 k_1.
\end{aligned}
\end{equation}

From \eqref{fX decrease bound}, we have
\begin{equation}\label{further fX bound}
\mathbb{E}\left[f\left(\alpha_{q+1}^l, \mathbf{X}_{q+1}^l\right) \mid \mathcal{F}_{q+1}^l\right] \leq f\left(\alpha_q^l, \mathbf{X}_q^l\right)+\left(t_q^l\right)^2 k_1.
\end{equation}

Relabel the sequence  $\left\{\left(\alpha_0^1, \mathbf{X}_0^1\right), \left(\alpha_1^1, \mathbf{X}_1^1\right), \ldots, \left(\alpha_{m_1-1}^1, \right. \right.$
$\left. \left. \mathbf{X}_{m_1}^1\!\right)\! ,\!\left(\alpha_1^2, \mathbf{X}_1^2\right), \left(\alpha_2^2, \mathbf{X}_2^2\right)\!, \!\ldots\!,\! \left(\alpha_{m_2}^2, \mathbf{X}_{m_2}^2\right)\!,\! \ldots\!,\! \left(\alpha_q^l, \mathbf{X}_q^l\right)\!,\! \ldots\right\}$ as $\left\{\left(\alpha_1, \mathbf{X}_1\right), \left(\alpha_2, \mathbf{X}_2\right), \ldots, \left(\alpha_{q-1}, \mathbf{X}_{q_1}\right), \ldots\right\}$. Similarly, we relabel $\left\{t_q^l\right\}$ and $\mathcal{F}_{q+1}^l$. According to \eqref{further fX bound} and \eqref{stepsize condition 1}, we obtain $\mathbb{E}\left[\left|f\left(\alpha_{q_1}, \mathbf{X}_{q_1}\right)+\sum_{l_2=q_1}^{\infty}\left(t_{l_2}\right)^2 k_1\right|\right]<\infty$. According to \cite{stochastic-processes}, we can conclude that $\left\{f\left(\alpha_{q_1}, \mathbf{X}_{q_1}\right)+\sum_{l_2=q_1}^{\infty}\left(t_{l_2}\right)^2 k_1\right\}$ is a non-negative supermartingale.

According to the martingale convergence theorem in \cite{super-martingale}, there exists a finite limit with probability 1, denoted as  $\lim _{q_1 \rightarrow \infty} \left[ f\left(\alpha_{q_1}, \mathbf{X}_{q_1}\right)+\sum_{l_2=q_1}^{\infty}\left(t_{l_2}\right)^2 k_1 \right]$, implying that $\lim _{q_1 \rightarrow \infty} f\left(\alpha_{q_1}, \mathbf{X}_{q_1}\right)$ exists, $\left\{f\left(\alpha_q^l, \mathbf{X}_q^l\right)\right\}$ converges.

Before probing that the proposed UM-SVRG algorithm converges to a stationary point, we present the following theorem.

\begin{theorem}\label{theorem-quasi-martingale}
 Let $\{X_n\}_{n \in \mathbb{N}}$ be a non-negative stochastic process with bounded positive variations, meaning it satisfies: $\sum_{n=0}^{\infty} \mathbb{E}\left[\mathbb{E}\left[X_{n+1}-X_n \mid \mathcal{F}_n\right]^{+}\right]<\infty$, where $X^{+}$ denotes the positive part of a random variable $X$, defined as $\max \{X, 0\}$, and $\mathcal{F}_n$ represents the increasing sequence of $\sigma$-algebras generated before time $n$. In this case, the stochastic process $\{X_n\}_{n \in \mathbb{N}}$ is a quasi martingale\cite{quasi-martingales}, i.e.,
\begin{equation}\label{quasi-martingale}
\sum_{n=0}^{\infty}\left|\mathbb{E}\left[X_{n+1}-X_n \mid \mathcal{F}_n\right]\right|<\infty,
 \end{equation}
and $ X_n$ converges a.s.
\end{theorem}

We accumulate inequality \eqref{fX decrease bound} from the current $q, l$ to $l \rightarrow \infty$ as follows
\begin{equation}\label{step grad inequation}
\begin{aligned}
&\sum_{l, q} t_q^l\left\|\operatorname{grad} f\left(\alpha_q^l, \mathbf{X}_q^l\right)\right\|_{\left(\alpha_q^l, \mathbf{X}_q^l\right)}^2 \leq \\
&-\sum_{l, q} \mathbb{E}\!\left[f\left(\alpha_{q+1}^l, \mathbf{X}_{q+1}^l\right)\!-\!f\left(\alpha_q^l, \mathbf{X}_q^l\right)\!\mid\!\mathcal{F}_{q+1}^l\right]\!+\!\sum_{l, q}\left(t_q^l\right)^2 k_1,
\end{aligned}
\end{equation}

From \eqref{fX decrease bound}, we have
\begin{equation}\label{step grad bound}
\sum_{l, q} \mathbb{E}\!\left[\!\mathbb{E}\!\left[f\!\left(\alpha_{q+1}^l, \mathbf{X}_{q+1}^l\right)\!-\!f\!\left(\alpha_q^l, \mathbf{X}_q^l\right)\!\mid\! \mathcal{F}_{q+1}^l\right]^{+}\!\right]\!\leq\! \sum_{l, q}\left(t_q^l\right)^2 k_1 .
\end{equation}

According to Theorem \ref{theorem-quasi-martingale} we have
\begin{equation}\label{absolute value bound}
\begin{aligned}
&\left|-\sum_{l, q} \mathbb{E}\left[f\left(\alpha_{q+1}^l, \mathbf{X}_{q+1}^l\right)-f\left(\alpha_q^l, \mathbf{X}_q^l\right) \mid \mathcal{F}_{q+1}^l\right]\right|\\
 &\leq \!\sum_{l, q}\left|\mathbb{E}\left[f\left(\alpha_{q+1}^l, \mathbf{X}_{q+1}^l\right)\!-\!f\left(\alpha_q^l, \mathbf{X}_q^l\right) \mid \mathcal{F}_{q+1}^l\right]\right| \!<\! \infty,
\end{aligned}
\end{equation}

Therefore, we have
\begin{equation}\label{step grad convergence}
\sum_{l, q} t_q^l\left\|\operatorname{grad} f\left(\alpha_q^l, \mathbf{X}_q^l\right)\right\|_{\left(\alpha_q^l, \mathbf{X}_q^l\right)}^2<\infty,
\end{equation}
and $\sum_{l, q} t_q^l\left\|\operatorname{grad} f\left(\alpha_1^l, \mathbf{X}_q^l\right)\right\|_{\left(\alpha_q^l, \mathbf{X}_q^l\right)}^2$ converges a.s. Additionally, $\lim _{l \rightarrow \infty} t_q^l\left\|\operatorname{grad} f\left(\alpha_q^l, \mathbf{X}_q^l\right)\right\|_{\left(\alpha_1^l, \mathbf{x}_q^l\right)}^2=0$. Then, we prove the convergence of $\|\operatorname{grad} f\left(\alpha_q^l, \mathbf{X}_q^l\right)\|_{\left(\alpha_q^l, \mathbf{X}_q^l\right)}^2$.

Here, following reference \cite{R-SGD}, we analyze the non-negative stochastic process $p_q^l=\|\operatorname{grad} f\left(\alpha_q^l, \mathbf{X}_q^l\right)\|_{\left(\alpha_q^l, \mathbf{X}_q^l\right)}^2$. Using Taylor expansion we have

\begin{equation}\label{gradient decrease bound}
\begin{aligned}
\mathbb{E}\left[p_{q+1}^l-p_q^l \mid \mathcal{F}_{q+1}^l\right] &\leq 2 t_q^l\left\|\operatorname{grad} f\left(\alpha_q^l, \mathbf{X}_q^l\right)\right\|_{\left(\alpha_q^l, \mathbf{X}_q^l\right)}^2 k_3\\
&+\left(t_q^l\right)^2 C k_2,
\end{aligned}
\end{equation}
where $k_2$ is the largest eigenvalue of the Hessian of $\|\operatorname{grad} f\left(\alpha_q^l, \mathbf{X}_q^l\right)\|_{\left(\alpha_q^l, \mathbf{X}_q^l\right)}^2$, and $-k_3$ is the lower bound of the minimum eigenvalue of the Hessian matrix of $f$ on $\mathcal{M}$.

Then, we can get
\begin{equation}\label{gradient convergence}
\sum_{l, q} \mathbb{E}\left[\mathbb{E}\left[p_{q+1}^l-p_q^l \mid \mathcal{F}_{q+1}^l\right]^{+}\right]<\infty,
\end{equation}
and $\{p_q^l\}$  is a quasi martingale. According to Theorem \ref{theorem-quasi-martingale}, $\{p_q^l\}$ converges. According to equations \eqref{stepsize condition 1} and \eqref{stepsize condition 2}, $\lim _{l \rightarrow \infty} t_q^l \rightarrow 0$ but not equal to 0. Therefore, $\{p_q^l\}$ converges to 0, indicating that the algorithm converges to a stationary point. This concludes the proof. $\hfill\blacksquare$

\ifCLASSOPTIONcaptionsoff
  \newpage
\fi

\end{document}